**From Plastic Waste to Treasure: Selective Upcycling through Catalytic Technologies**

*Shuai Yue, Pengfei Wang*, Bingnan Yu, Tao Zhang, Zhiyong Zhao, Yi Li and Sihui Zhan**

S. Yue, Prof. P. Wang, B. Yu, T. Zhang, Z. Zhao, Prof. S. Zhan
MOE Key Laboratory of Pollution Processes and Environmental Criteria Tianjin Key Laboratory of Environmental Remediation and Pollution Control
College of Environmental Science and Engineering
Nankai University
Tianjin 300350, P. R. China
E-mail: sihuizhan@nankai.edu.cn; pengfeiwang@hebut.edu.cn
Prof. Y. Li
Tianjin Key Laboratory of Molecular Optoelectronic Sciences
Department of Chemistry, School of Science
Collaborative Innovation Center of Chemical Science and Engineering (Tianjin)
Tianjin University
Tianjin 300072, P. R. China

Keywords: (Plastic waste, Selective upcycling, Catalytic technologies, Reaction mechanisms, Circular economy)

(Abstract: The huge amount of plastic wastes has become a pressing global environmental problem, leading to severe environmental pollution and resource depletion through conventional downcycling technologies like incineration and landfilling. In contrast, selective upcycling of various plastics offers a promising solution for converting waste plastics into valuable products. This review provides a comprehensive overview of the recent advancements in innovative catalytic technologies, including thermocatalysis, electrocatalysis, and photocatalysis. Special emphasis is placed on elucidating the reaction mechanisms, activating designated chemical bonds for high selectivity, and elaborating the above techniques in terms of reaction conditions and products. Finally, the application prospects and future development trends in plastic catalysis are discussed, providing valuable insights for realizing a sustainable circular plastic economy.)

**1. Introduction**

Plastics are widely used in daily life because of their practicality, low cost, and stable chemical properties.[1-3] However, the rapid accumulation of plastic waste is a pressing global environmental problem.[4-5] By 2050, it is estimated that the amount of plastic waste in nature will reach 1.2 billion tons, with more than 80% being directly discarded or landfilled.[6-7] Due to the chemical inertia of plastics, natural degradation can take hundreds of years. Moreover, the degradation process produces a large amount of microplastics (defined as plastics with diameter ≤5 mm).[8-10] These microplastics can enter the animal and human food chain and pose a serious threat to the ecosystem stability.[11-12] Therefore, there is an urgent need to develop new strategies for the efficient treatment and recycling of plastics to address this global environmental problem.[13]

At present, plastic treatment and conversion methods include traditional methods such as landfilling, incineration, and mechanical recycling, as well as newer catalytic technologies such as thermocatalysis, electrocatalysis, and photocatalysis (**Figure 1**). **Table 1** summarizes the existing plastic conversion technologies. Direct landfilling or incineration of plastic waste is the most commonly used method. However, it generates harmful substances such as dioxins and contributes to greenhouse gas emissions, which does not comply with the low-carbon environmental protection strategies.[6] Mechanical recycling—another commonly used method, involves physical processes such as extrusion, grinding, and crushing, which do not alter the chemical composition of plastics but can significantly degrade their mechanical properties after multiple cycles, adding limited value in terms of recycling.[14] Moreover, the limited number of recycles and the range of application hinder the widespread application of this technology.[15]

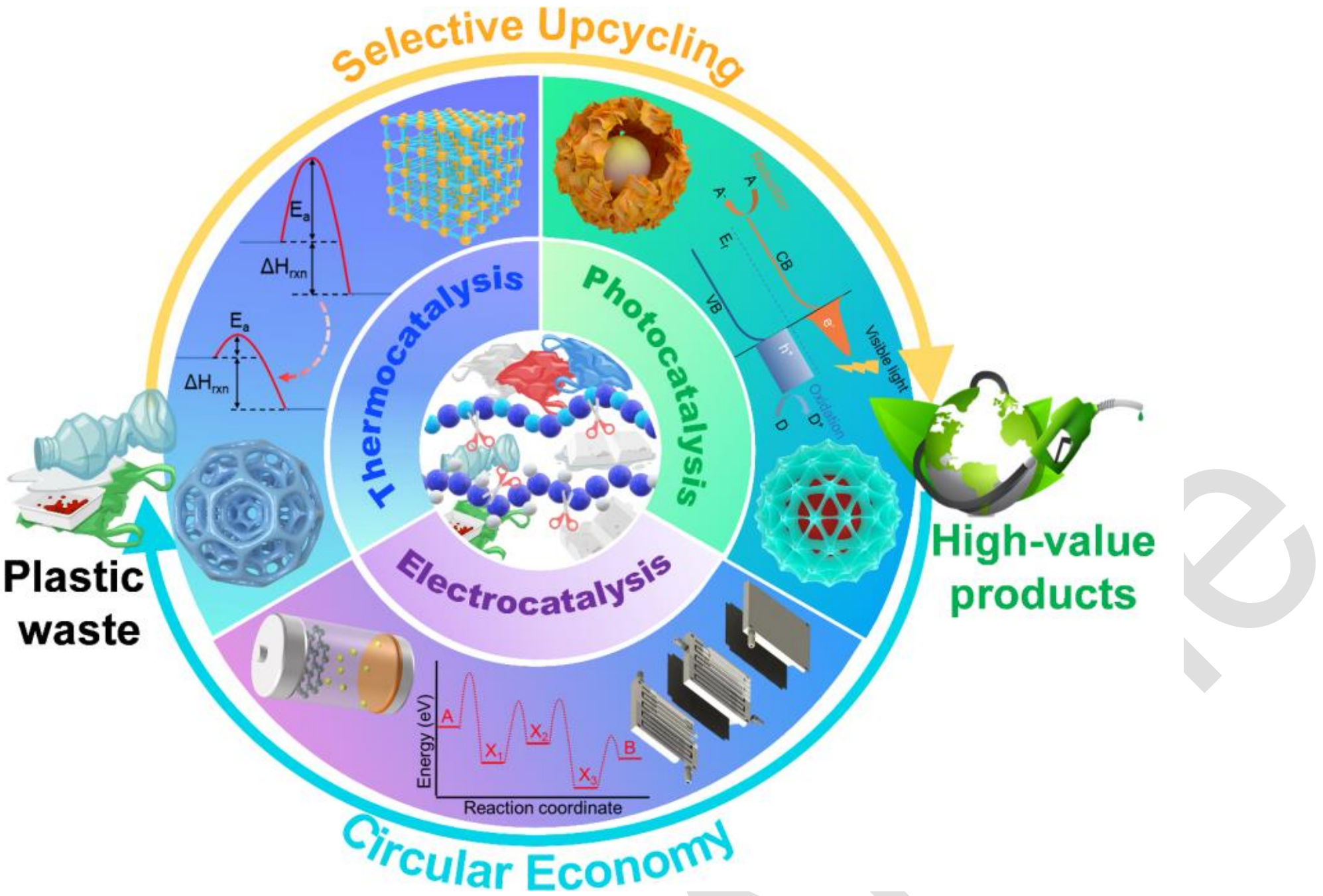

**Figure 1.** Schematic diagram of newer catalytic technologies for plastic upcycling.

**Table 1.** Strategies for plastic waste treatment.

| Categories | Techniques | Advantages | Disadvantages |
|---|---|---|---|
| Upcycling | Photocatalysis | Produce high-value products<br>Mild reaction conditions<br>High product selectivity<br>Use solar energy | Low reaction efficiency<br>Low reaction stability |
| | Thermocatalysis | Produce high-value products<br>High reaction efficiency<br>High reaction stability | Strict reaction conditions<br>High energy consumption |
| | Electrocatalysis | Produce high-value products<br>Mild reaction conditions<br>High reaction efficiency | Limited application scope<br>High energy consumption |
| Downcycling | Mechanical recycling | Produce regenerative plastics | Strict pretreatment<br>Raw materials limitations |
| No recycling | Incineration | Feedstock versatility<br>Electricity production | Release toxic compounds<br>Massive $CO_2$ emissions |
| | Landfill | Simple and low cost | Unsustainability<br>Soil and groundwater pollution |

In contrast to traditional plastic recycling strategies, the upcycling of plastics into high-value chemicals and green fuels using catalytic technologies is attracting increasing research attention (**Figure 2**).[16-18] Although thermocatalytic conversion requires higher temperatures, it ensures high reaction efficiencies and can significantly suppress side reactions.[19] Similarly, electrocatalytic conversion for plastic reforming requires additional electrical energy input. However, this technique ensures good selectivity, an impressive reaction rate, and more efficient use of electrical energy to achieve precise conversion of C-C bonds, accompanied by the production of green fuels such as $H_2$ and $CH_4$.[20] Photocatalytic plastic upcycling relies only on solar energy, and the reaction conditions are milder. Although it can achieve precise activation of specific chemical bonds and the conversion of waste plastics into high-value chemicals, its reaction efficiency needs to be improved.[21]

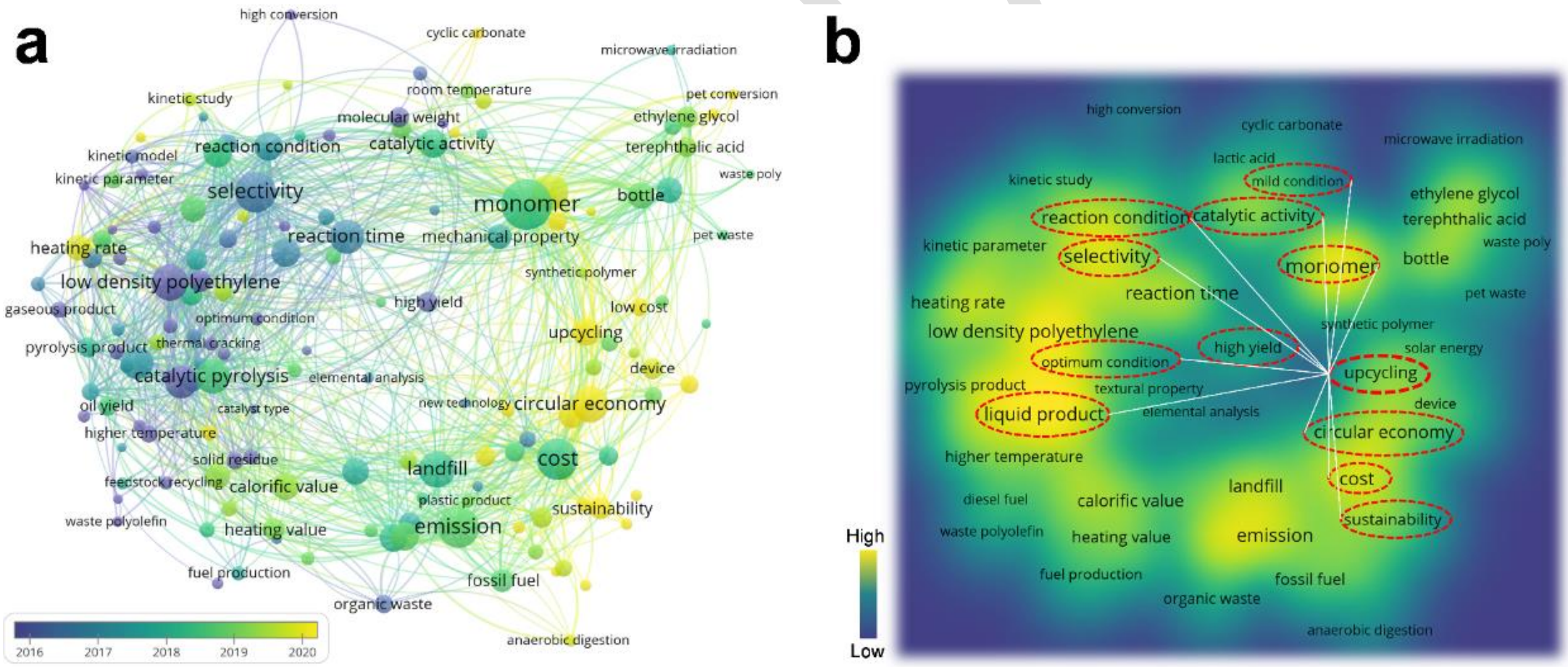

**Figure 2.** (a) Keyword clouds were generated to provide a correlation between most researched processes and hotspots along with time. It was created by text-mining published plastic recycling articles from 2016 to the present. (b) Corresponding keyword thermodynamic diagram. Red circles show the keywords related to upcycling.

Catalytic conversion of plastics can be categorized into two types: indirect and direct upcycling. In indirect upcycling routes, plastics are initially depolymerized into monomers, oligomers, or derivatives, which are subsequently converted into high-value chemicals containing specific functional groups. The microstructure of the polymers plays a critical role in selective depolymerization, which involves pathways mediated by monomers, oligomers, or derivatives.[17] As shown in **Figure 3a**, the process involves polymers that contain ester or amide linkages in their main chains. Treatment with alcohols, ammonia (amines), bases, and other reagents (represented by Y-H) facilitates the decomposition of plastics into monomers or derivatives, which are further subjected to catalytic conversion into high-value products. For plastic polymers with C-C bonds, the depolymerized monomers can introduce specific chemical functional groups to replace the hydrogen atoms in monomers and oligomers, enabling their selective transformation into heteroatom-containing chemicals (**Figure 3b**). Additionally, plastics can be directly degraded into $C_1$ module molecules (platform molecules) such as $CO_2$ and $CH_4$ using catalysts.[22] The conversion of these platform molecules into high-value carbon ($C_{2+}$) fuels or other value-added chemicals has attracted considerable attention (**Figure 3c**).

In addition to indirect conversion, the direct conversion of plastics is a strategy with great potential.[23] As a macromolecular structure, plastic comprises a large number of repeating monomer units connected by specific types of chemical bonds.[24] The precise activation of these chemical bonds forms the basis for the design of catalytic systems. As illustrated in **Figure 3d**, direct hydrogenolysis of plastics is a powerful approach for polymer conversion because the monomer units in polymers are readily reduced and cleaved by hydrogen,[25] thereby enabling the conversion of plastic wastes into high-value chemicals. Direct conversion can also be achieved through coupled exothermic hydrogenation and endothermic aromatization reactions, suggesting the feasibility of in situ hydrogen species generation (e.g., self-sustaining hydrogenolysis of polymers or the decomposition of another hydrogen donor) to drive the process selectively (**Figure 3e**).[26] Moreover, during the depolymerization of plastics, heteroatoms such as O, N, and S are directly introduced to attack the C-C/C-H bonds, facilitating the incorporation of high-value functional groups (**Figure 3f**).[27] Subsequently, catalysts have been utilized to accomplish the depolymerization and functionalization of plastics, representing an ideal pathway for the selective catalytic conversion of plastics.

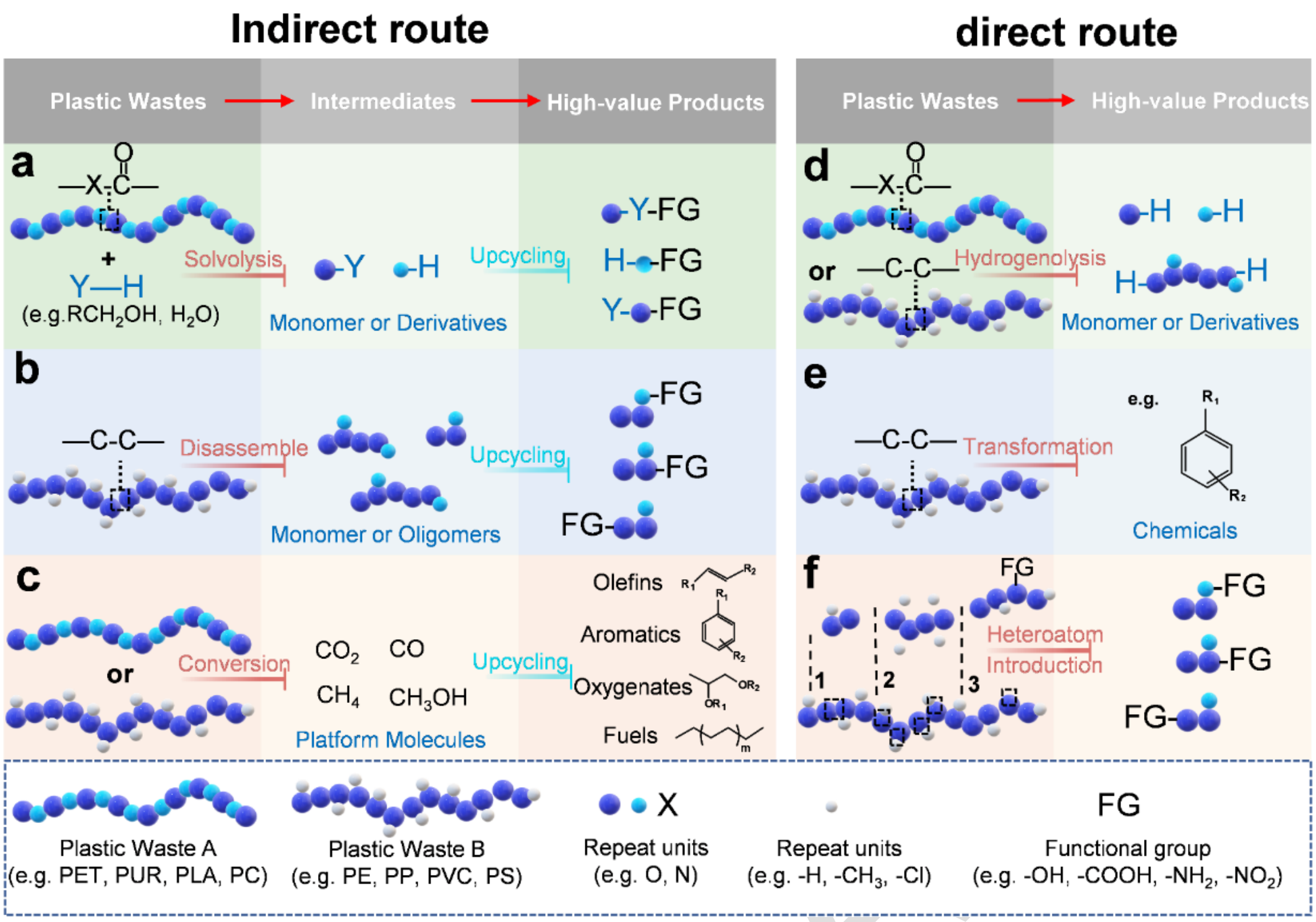

**Figure 3.** Schematic illustration of indirect and direct plastic upcycling routes. (a, b) The sub-type of monomer-, oligomer-, or derivate-mediated routes. (c) The route mediated by small platform molecules. (d) The directly disassembled routes via the activation and cleavage of specific chemical bonds without successive transformations. (e) The direct deconstruction routes undergo further transformation to chemicals with higher value. (f) The functionalized routes by inserting chemical functional groups.

In recent years, numerous reviews have focused on the degradation or recycling of plastic wastes, primarily outlining the strategies for converting plastic wastes into chemicals.[28-29] However, attention has been rarely paid to the internal conversion mechanism and selective conversion process of plastic reforming. Researchers have primarily focused on individual catalytic technologies, neglecting a comprehensive study of the novel catalytic technologies like thermocatalysis, electrocatalysis, and photocatalysis. In this regard, a comprehensive overview of the existing catalytic plastic upcycling technologies can provide readers with an in-depth understanding of the advancements in plastic reforming technologies.

## 2. Thermocatalytic conversion of plastics

Thermocatalytic conversion of plastics occurs at a certain temperature and pressure. Methods that have been explored in this regard include pyrolytic conversion,[30] catalytic hydrogenation,[31] and field-assisted thermal conversion.[32] In the context of thermocatalytic plastic conversion, we put particular emphasis on the selective upcycling of waste plastics into high-value products such as liquid fuels and hydrocarbons, highlighting the recent advancements in the field and assessing their potential to create a circular economy for plastics.

**2.1. Thermocatalytic conversion process and mechanism**

Thermocatalytic conversion primarily involves overcoming the thermodynamic energy barrier in a catalytic reaction system via heating, thereby promoting the conversion of reactants to products.[33] Therefore, it is essential to consider the deconstruction and upcycling of plastics from thermodynamic and kinetic perspectives. As depicted in **Figure 4**, the selective conversion of polymers to the desired products involves a certain reaction barrier ($E_a$) and an enthalpy difference ($\Delta H_{rxn}$) between the reactants and products.[17] These factors determine the favorability of the free energy of the reaction. Because the entropy change ($\Delta S_{rxn}$) during the depolymerization of plastics is positive and $\Delta H_{rxn}$ varies significantly depending on the target product, catalysts can effectively reduce $E_a$ during thermocatalytic reactions while leaving $\Delta H_{rxn}$ unchanged.[34] This indicates that catalysts do not alter the thermodynamics of the reaction but can modify the reaction rates through product substitution mechanisms. The reaction temperature required for plastic depolymerization plays a crucial role and is primarily influenced by the upper temperature limit ($T_c$) of the polymer.[35] Therefore, optimization of the catalysts and reaction conditions is a primary focus in thermocatalytic conversions.

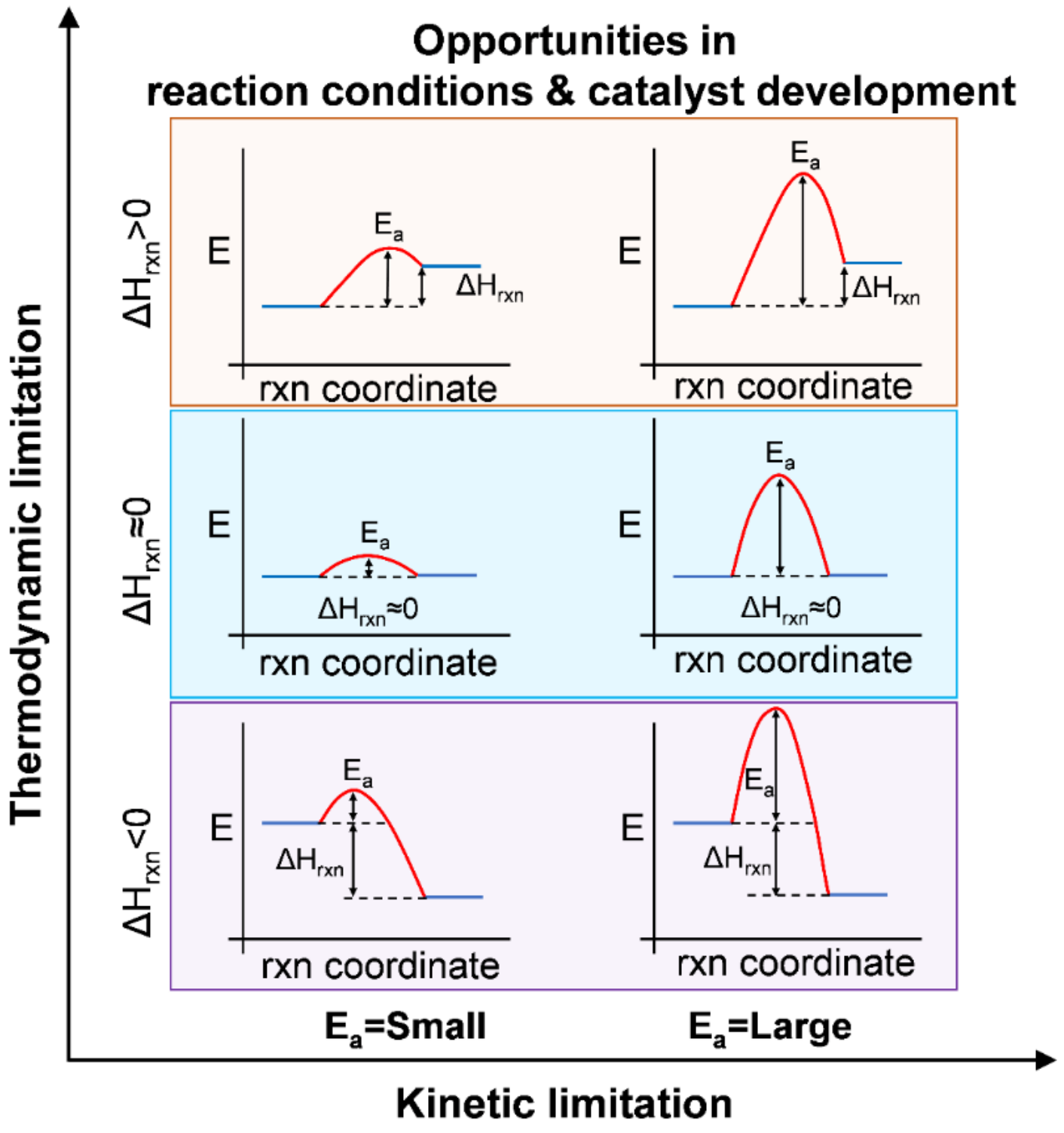

**Figure 4.** Schematic illustration of thermodynamic and kinetic control in plastic upcycling. It is classified as $\Delta H_{rxn}>0$, $\Delta H_{rxn}\approx0$, $\Delta H_{rxn}<0$ from the thermodynamic limitation, corresponding to the kinetic limitation $E_a$=Small or $E_a$=Large.

## 2.2. Overview of Recent Advances and Studies

### *2.2.1. Thermocatalytic conversion under extreme conditions*

Pyrolysis can convert plastic polymers into high-value chemicals such as fuel oils under extreme conditions of high temperature and pressure, achieving a high recovery rate and allowing for the rapid cleavage of plastic chains to produce valuable chemicals.[36] However, the complexity of the pyrolysis process, which includes terminal chain depolymerization, random chain depolymerization, side chain elimination, and monomer formation, renders it challenging to fully comprehend and summarize the reaction mechanisms.[37] The lack of selectivity in high-temperature pyrolysis processes also leads to highly complex product compositions, limiting the secondary utilization of plastic conversion products (**Table 2**). Therefore, efficient and selective pyrolysis of plastics are critical for future advancements in this field.

**Table 2.** Different pyrolysis technologies and corresponding products.

| Technologies | Plastics | Conditions | Products | Refs. |
|---|---|---|---|---|
| Thermal pyrolysis | LDPE | 793 K | $C_5$–$C_{15}$ hydrocarbons and $C_{12}$–$C_{28}$ light oil | [38] |
| | HDPE | 723 K and 773 K | Paraffin, olefin, naphthene and aromatic | [39] |
| | PP | 793 K | $C_5$–$C_{15}$ hydrocarbons and $C_{12}$–$C_{28}$ light oil | [38] |
| Catalytic pyrolysis | LDPE | 825 K | Monoaromatics and $C_5$–$C_{12}$ alkanes/olefins | [40] |
| | HDPE | 723–743 K | $C_{13}$–$C_{23+}$ paraffin | [41] |
| | PS | 673 K | $C_9$–$C_{12}$ alkylaromatic, methylstyrene and isopropylbenzene | [42] |
| Hydropyrolysis | PP | 523 K | Gasoline ($C_1$–$C_{12}$) and diesel ($C_9$–$C_{22}$) | [43] |
| | PET | 923 K | Paraffins, olefins and aromatics | [44] |

Considering the above, Scott et al.[45] conducted a tandem catalytic reaction using Pt/γ-$Al_2O_3$ (**Figure 5a**), achieving high yields (up to 80 wt%) in the conversion of various polyethylene grades. Despite a high reaction temperature of 280 °C, this method demonstrated excellent selectivity, producing high-value long-chain alkylaromatics and alkylnaphthalenes as the main reaction products. Gel permeation chromatography with refractive index detection (GPC-RI) (**Figure 5b**) revealed a tenfold reduction in the molecular weight ($M_w$) of polyethylene. The temporal evolution of the reaction products (**Figure 5c**) revealed a significant increase in the fraction of liquid hydrocarbons during a short induction period. To investigate the thermocatalytic reforming mechanism of PE (**Figure 5d-f**), the reaction products were examined by NMR spectroscopy, which revealed that most of the aromatic protons were bound to the benzene ring; no alkene or diene was detected during the reaction. This implies that the polyethylene chain can absorb the hydrogen produced during aromatization, thereby promoting the aromatization reaction. The Pt/γ-$Al_2O_3$-catalyzed tandem hydrogenolysis/aromatization reaction of polyethylene enables the direct conversion of waste polyethylene plastics into high-value alkylaromatics and alkylcycloalkanes. This process effectively converted polyethylene into low-molecular-weight liquid/wax products, eliminating the need for additional solvents and hydrogen. Notably, a relatively high reaction temperature, reaching 280 °C, is required in this method. Consequently, the current application cost of this approach remains high, limiting its practical implementation.

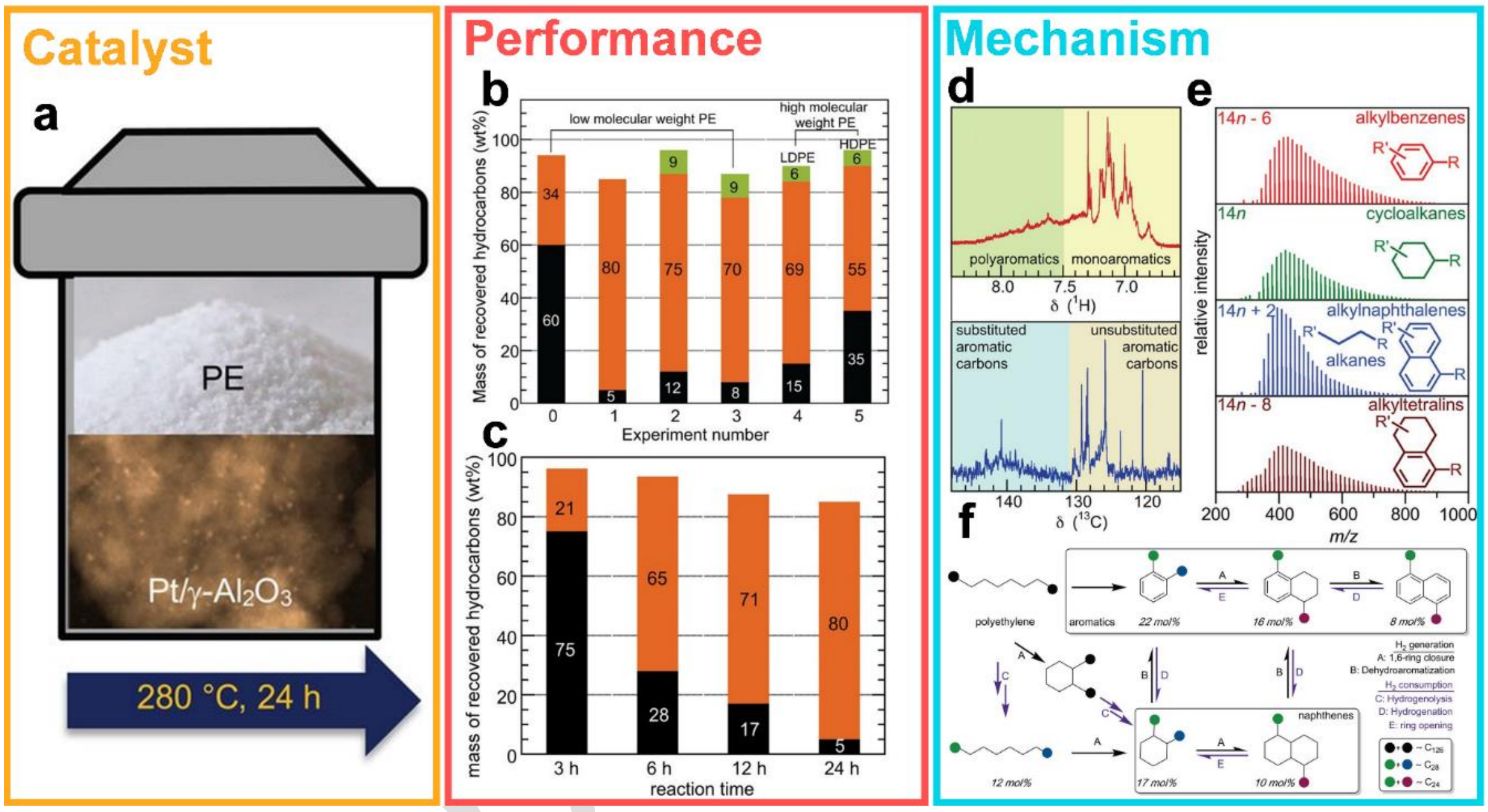

**Figure 5.** (a) Schematic of reactor with photographs of the powdered polymer and a transmission electron micrograph of the catalyst. (b) Hydrocarbon distributions after 24 hours at 280°C. (c) Evolution of major product fractions catalyzed by Pt/γ-$Al_2O_3$ in an unstirred mini-autoclave reactor at 280°C. (d) $^1$H and $^{13}$C NMR spectra, recorded in deuterated TCE. (e) FD-MS analysis in the solvent-free catalytic conversion of polyethylene. (f) Overall PE conversion to alkylaromatics and alkylnaphthenes, and proposed mechanism of tandem polyethylene hydrogenolysis/aromatization via dehydrocyclization. Reproduced with permission.[45] Copyright 2020, AAAS.

The highly selective upcycling of waste plastics into high-value chemicals is a critical step toward achieving a global circular economy. Ma et al.[46] introduced a novel two-step heterogeneous catalytic system (**Figure 6a**) to convert PLA plastics into methyl methacrylate. Using an α-MoC catalyst, methyl propionate could be efficiently produced from PLA waste plastics through alcoholysis and hydrodeoxygenation (**Figure 6b**). The conversion of actual plastic materials, such as straws, cups, and tableware, to methyl propionate was also attempted, achieving the complete conversion of polylactic acid and a methyl propionate (MP) yield of 98% (**Figure 6c**). Analysis of the reaction process indicated that polylactic acid underwent alcoholysis to generate methyl lactate, which was subsequently converted to MP via catalytic hydrodeoxygenation (**Figure 6d,e**). Furthermore, Cs-La/$SiO_2$ was used as a catalyst for the conversion of PLA to produce both methyl methacrylate and MP (**Figure 6f**). Thus, gram-level polylactic acid plastics could be successfully converted to methyl methacrylate through a two-step catalytic process, as summarized in **Figure 6g-i**. This innovative method for the highly selective conversion of PLA into high-value chemicals is advantageous over natural degradation pathways. Compared to the traditional monomer recovery process, this route offers a novel recycling strategy for the production of high-value chemicals with high conversion and selectivity. However, an elevated reaction temperature necessitates relatively high processing costs, hampering its widespread practical application. Additionally, the two-step reaction process renders the reaction rate more dependent on various conditions.

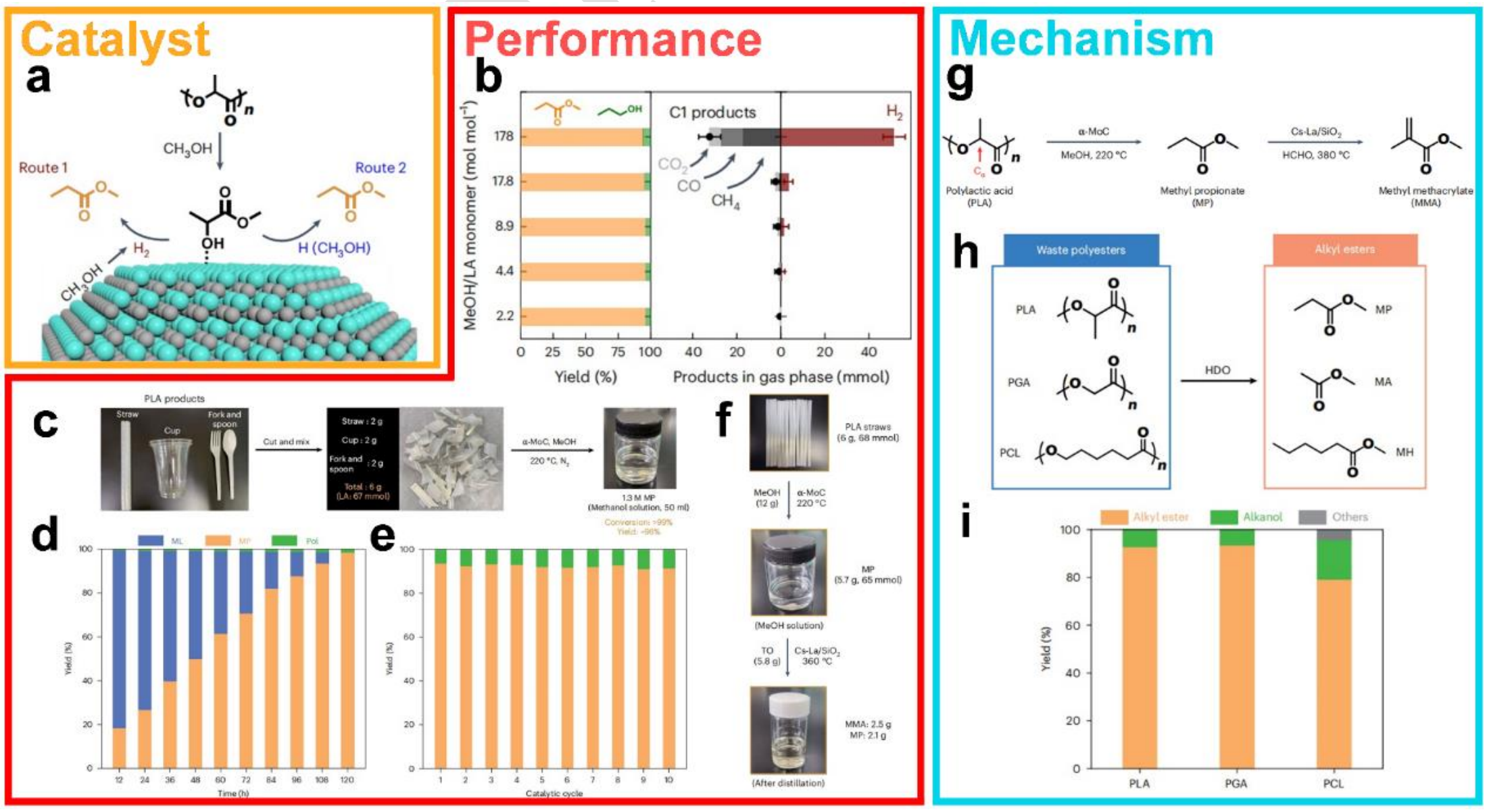

**Figure 6.** (a) Schematic reaction pathways for selective HDO of PLA on catalyst. (b) The product distribution from HDO of PLA with different molar ratios of MeOH/LA monomer. Schematic process (c) and time-dependent conversion (d) of commercial PLA products towards MP. (e) Repeated catalytic cycles of PLA degradation. (f) Schematic pathway of MMA production from waste PLA plastics through two-step catalytic process.(g) Upcycling route

from PLA to MMA. Chemical structures (h) and product yields (i) of different polyesters converted to alkyl esters through HDO. Reproduced with permission.[46] Copyright 2023, Springer Nature.

*2.2.2. Thermocatalytic conversion under mild conditions*

The pursuit of mild reaction conditions in pyrolytic conversion is essential for realizing low-cost plastic conversion. High reaction temperatures are often required for a series of catalytic reactions, limiting their practical industrial applications.[47] Lercher et al.[48] proposed the utilization of highly ionic reactions to catalyze plastic polymers and reduce the energy of the ionic transition states, enabling the complete conversion of polyethylene and polypropylene into liquid isoalkanes at temperatures below 100 °C. The experiments were conducted using commercial low-density polyethylene (LDPE) and $iC_5$ as substrates in a glass-tube reactor, as depicted in **Figure 7a**. The presence of Lewis acidic chloroaluminate ions, along with $iC_5$, facilitated the complete conversion of LDPE after 3 h at 70 °C (**Figure 7b**). This strategy enables the near-quantitative and selective conversion of a wide range of polymeric consumer products to a total of more than 70 wt% $C_6$–$C_{12}$ products (**Figure 7c**). To account for the tandem reactions proceeding at a constant rate, this study assumed two carbocation-based ligated catalytic cycles with olefins derived via cracking as the ligated intermediates (**Figure 7d**). Carbocations formed in the polymer chains tend to undergo skeletal isomerization and cracking through beta-scission. The study introduced a transformative and scalable approach for the conversion of polyolefins. The mild reaction temperature employed in the process significantly reduces the cost of catalytic plastic conversion. However, contrary to the principles of environmental protection, the use of additional chemicals currently restricts their application prospects.

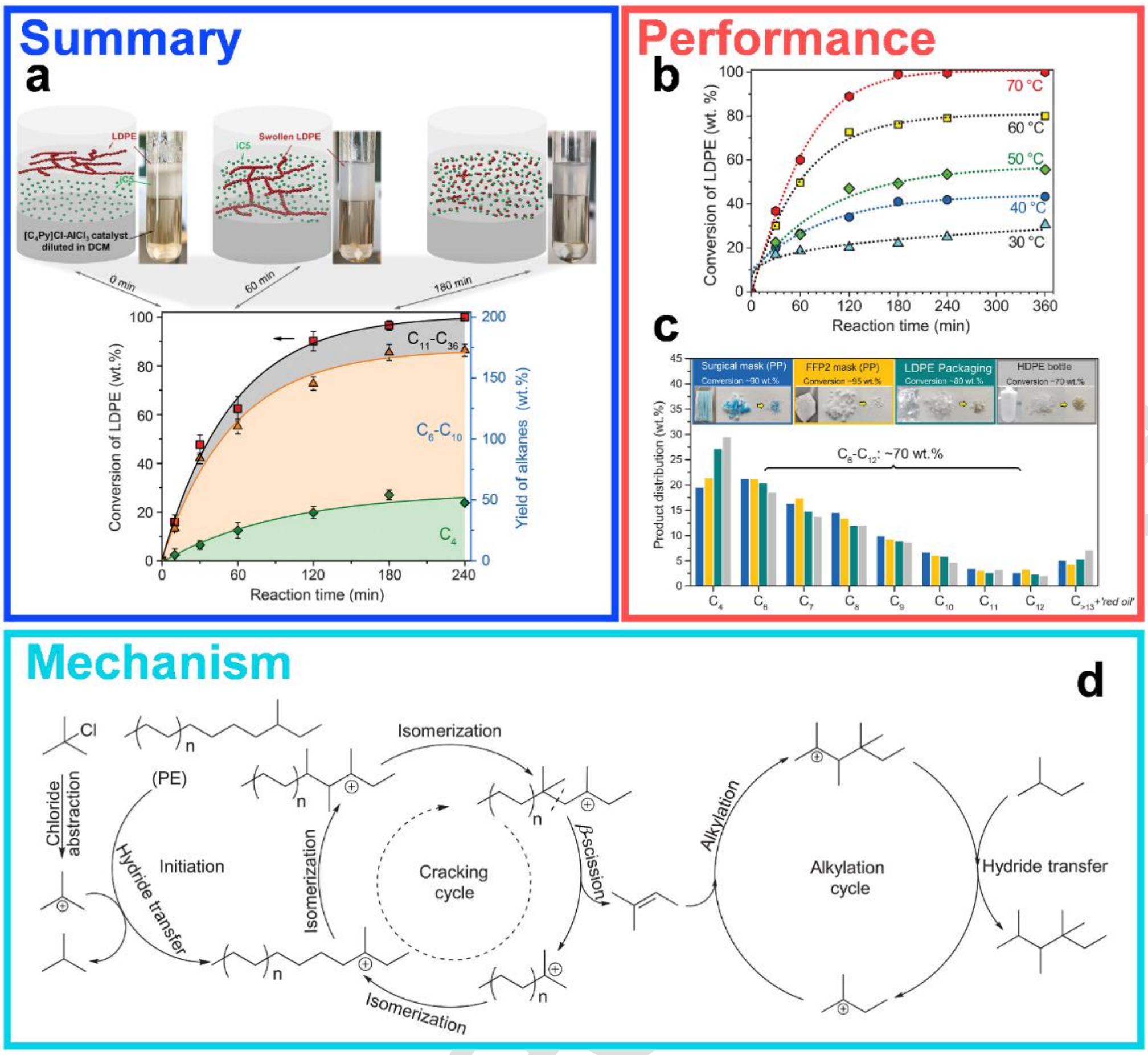

**Figure 7.** (a) One-pot catalytic LDPE and $iC_5$ upcycling into liquid alkanes over Lewis acidic chloroaluminate ionic liquid at 70°C. (b) Time-resolved conversion profile of LDPE in the presence of TBC at different temperatures. (c) Selective deconstruction of postconsumer polyolefin waste into liquid alkanes. (d) Proposed reaction mechanism for the tandem cracking-alkylation process of a polyolefin with $iC_5$. Reproduced with permission.[48] Copyright 2023, AAAS.

Currently, researchers are focusing on the use of catalysts to develop more efficient and environmentally friendly plastic recycling routes with milder reaction conditions.[49] Niu et al.[50] demonstrated the efficacy of a dinuclear composite catalyst for the degradation of polyethylene terephthalate (PET) and other widely used plastics in daily life. The binuclear zinc metal catalyst $Zn_2L(NO_3)_2$ was synthesized based on the concept of binuclear metallohydrolases (**Figure 8a**). The depolymerization kinetics of PET were evaluated under mild environmental conditions, and complete depolymerization into monomers was observed within 10 weeks (**Figure 8b**). The hydrolysis kinetics of $Zn_2L(NO_3)_2$ were comparable for crystalline PET particles (38%) and amorphous PET at pH 8 and 60 °C, indicating the unique capability of $Zn_2L(NO_3)_2$ to chemically decompose inert PET under mild conditions (**Figure 8c**). Furthermore, this study demonstrated the catalytic potential and cycle stability of $Zn_2L(NO_3)_2$ in the depolymerization of 13 different polyesters and nylons, highlighting its broad applicability (**Figure 8d**). DFT calculations revealed that the binuclear catalyst formed stable interactions with the EGD compound through two adjacent zinc sites (**Figure 8e**). This biomimetic binuclear Zn-site catalyst possesses intramolecular properties that enable the rapid hydrolysis of plastic polymers under mild conditions. However, the reaction products are relatively complex. To further advance plastic recycling, higher-value high-carbon products should be emphasized, as it aligns better with the establishment of a circular carbon economy.

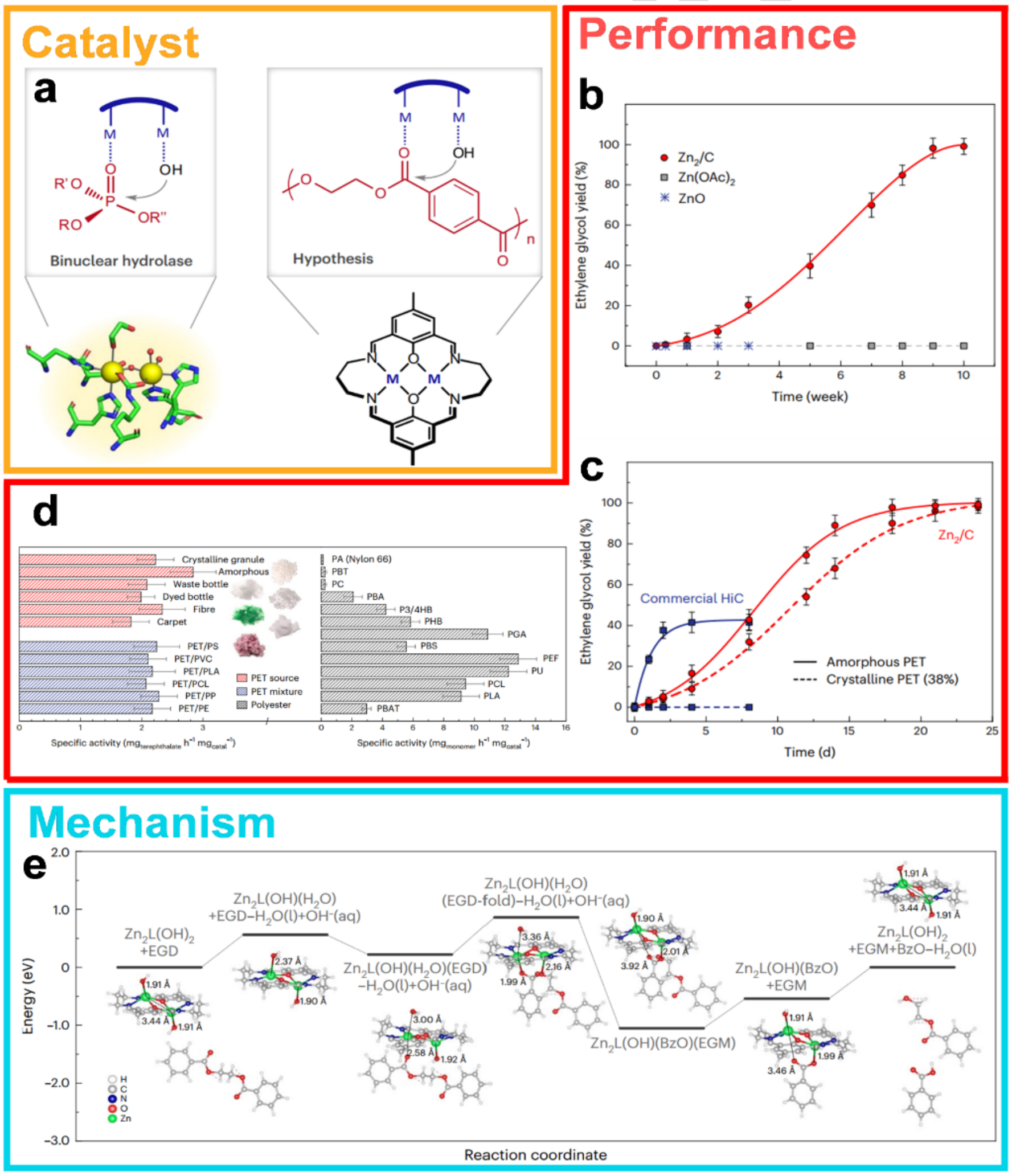

**Fig. 8** (a) Binuclear catalyst design. (b) The hydrolysis kinetics of am-PET over $Zn_2/C$, zinc acetate and zinc oxide at pH 8 and 40 °C. (c) The hydrolysis kinetics of am-PET and crystalline

PET granule (38%) over $Zn_2$/C and commercial HiC at pH 8 and 60 °C, an optimal condition for the commercial HiC. (d) The specific activities of $Zn_2$/C toward different PET sources, PET mixtures, and various polyesters and polyamide at pH 13 and 60 °C. Insets show the images of PET crystalline granules, amorphous flakes, clear and dyed bottle flakes, and cloth fibres. (e) Potential energy profile of EGD decomposition on $Zn_2L(OH)_2$ compound. The inset shows the DFT-optimized geometries for reactants and intermediates. Reproduced with permission.[50] Copyright 2023, Springer Nature.

### *2.2.3. Novel field-assisted thermocatalytic conversion under mild conditions*

As the name implies, field assistance involves the utilization of external fields, such as microwaves and piezoelectricity, to aid pyrolytic and catalytic processes, thereby facilitating the rapid conversion of plastics.[51,52] In microwave-assisted catalysis, the electromagnetic spectra encompass microwave frequencies ranging from 0.3 to 300 GHz.[53] Compared to conventional thermal catalysis, microwave-assisted catalysis ensures the uniform heating of plastics, resulting in high reaction rates and reduced energy consumption.[54]

Edwards et al.[32] introduced an innovative one-step microwave-assisted catalytic method for the rapid conversion of plastics. The method relied primarily on microwave-induced solid-solid catalytic reactions (**Figure 9a**). Under the influence of microwaves, the mechanical mixture of plastic and catalyst ($FeAlO_x$) underwent dissociation, resulting in the rapid release of a substantial amount of hydrogen gas, reaching a hydrogen yield of 55.6 mmol·$g^{-1}$. The remaining carbon residues predominantly existed in the form of carbon nanotubes (**Figure 9b,c**). This approach demonstrated a wide applicability to various plastics, including PP and PS (**Figure 9d**). Although microwave field-assisted catalysis has extensive applications, the mechanism governing microwave–compound interactions remains unclear.[55] The $FeAlO_x$ catalyst generated significant heat during microwave exposure, allowing to reach the required temperature difference for C-H bond cracking upon contact with the plastic substrate (**Figure 9e**). Consequently, intermediates, small olefins, and $H_2$ were formed, while carbon species infiltrated the iron catalyst lattice, leading to the formation of multi-walled carbon nanotubes (**Figure 9f**). The one-step microwave-induced catalytic bond cleavage of plastics demonstrated in this study offers an efficient and selective approach for transforming plastics, demonstrating its potential as a novel technique for practical engineering applications. However, carbon nanotubes obtained by microwave-assisted pyrolysis are still less attractive than high-value chemicals, and the microwave field input requires high power consumption, leading to high processing costs.

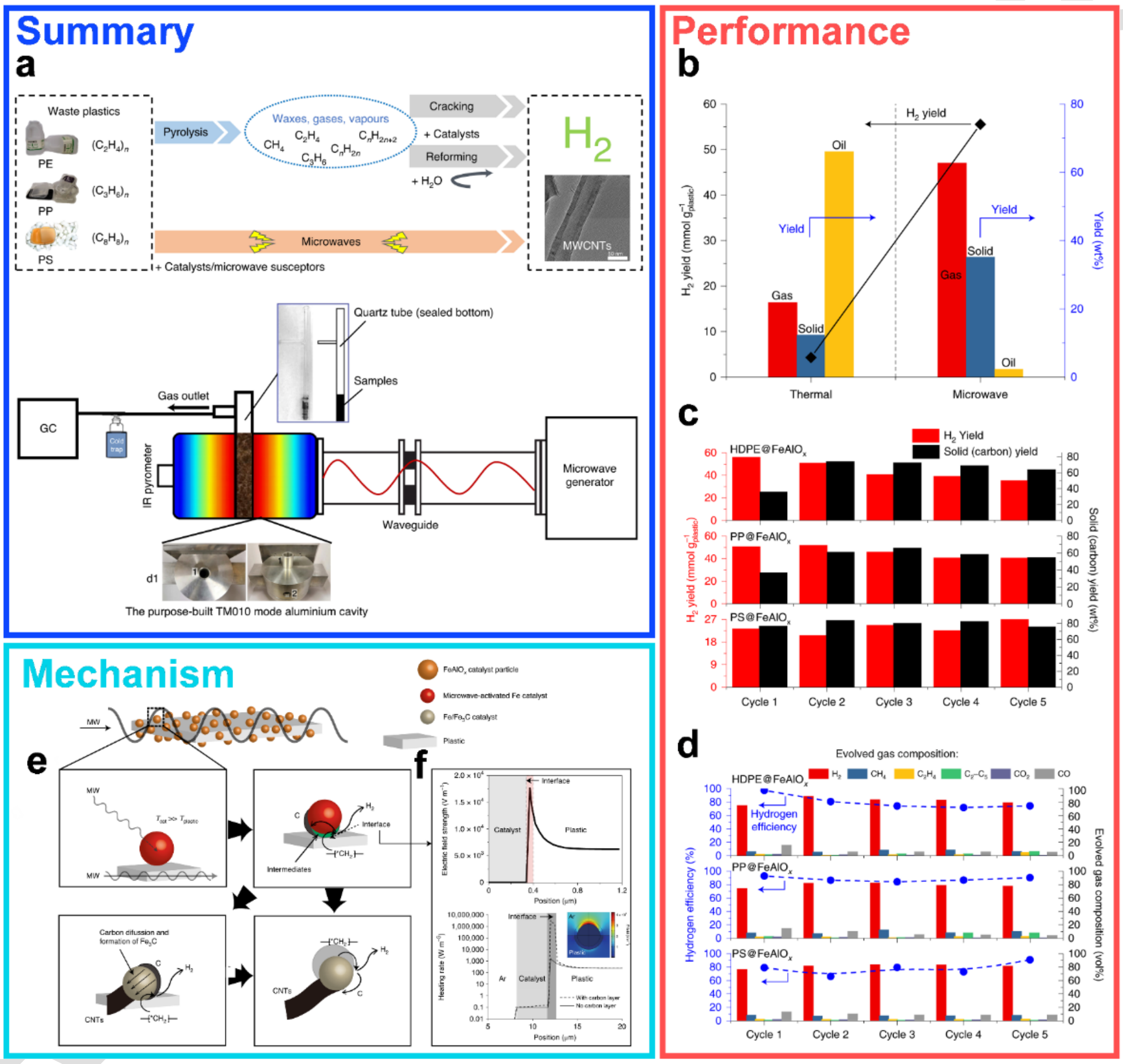

**Figure 9.** (a) The two-step pyrolysis and gasification process and the present one-step microwave-initiated catalytic process, and the corresponding experimental set-up and reaction system configuration. (b) A comparison of microwave and conventional thermal processes. (c) Successive cycles of the deconstruction of HDPE, polypropylene and polystyrene on $FeAlO_x$. (d) Hydrogen efficiency and evolved gas composition (vol%). (e) An illustration on microwave-initiated decomposition of plastic over a $FeAlO_x$ catalyst. (f) Modelling of the electric field strength and the heating rate at the interface. Reproduced with permission.[32] Copyright 2020, Springer Nature.

*2.2.4 Other notable catalytic plastic pyrolysis technologies*

In addition to the aforementioned plastic conversion technologies, the emerging catalytic pyrolysis technologies for plastic conversion are gaining attention because of the high selectivity and yield of high-value chemicals.[56] These technologies are at the forefront of research and development and are expected to drive future advancements in this field. For example, Hu et al.[57] proposed an innovative plastic reforming strategy based on electro-joule heating. This approach overcomes the challenges of poor selectivity and low yield of conventional plastic pyrolysis reactions. By utilizing electro-joule pulse heating technology, precise control of the reaction temperature in time and space can be achieved, leading to improved product selectivity and efficient conversion of plastics into monomers. This breakthrough opens up new possibilities for plastic recycling. **Table 3** summarizes the existing technologies to provide an overview of the current research advancements in the pyrolytic catalytic conversion of plastics. These approaches utilize high-energy-induced bond decomposition and high-temperature catalytic conversion to enhance the conversion efficiency. The primary focus is to achieve high selectivity for high-value chemicals and sustainable energy production. Consequently, the thermal catalytic conversion technology is advancing toward practical applications. However, maintaining a high and stable reaction temperature requires significant energy consumption, resulting in higher costs associated with this technology.

**Table 3.** Thermocatalytic conversion of plastics over different catalysts.

| Catalysts | Plastics | Conditions | Products | Selectivity | Refs. |
|---|---|---|---|---|---|
| L-$ZrO_2$@m$SiO_2$ | LDPE | 573 K, 1MPa | $C_{18}$-centred distribution | 39–54% | [58] |
| PDA-CNTs | PET | 423K | BHET | 50+% | [19] |
| ZnO zeolite | PET | 723 K | aromatic hydrocarbon | 83.9 % | [59] |
| Pt/$WO_3$/$ZrO_2$ mix HY zeolite | LDPE and PP | 523 K, 1.5 MPa | liquid fuels | 85% | [46] |
| sea shell-derived catalyst | PLA | 773 K | lactic acid | 12% | [60] |
| $SO_4^{2-}$–$ZrO_2$ | HDPE | 773-973 K | ethylene | 25% | [61] |
| Ru/C | PE | 473 K, 2 MP | liquid n-alkanes | 45% | [62] |
| Ru/$TiO_2$ | i-PP | 523 K, 3MPa | valuable lubricant-range hydrocarbons | 80+% | [63] |
| Ir | HDPE | 423 K | linear alkane | 56% | [64] |
| β-zeolite/Pt@silicalite | LDPE | 523 K | naphtha | 89.5% | [65] |

## 3. Electrocatalytic conversion of plastics

Electrocatalytic technologies enable the conversion of plastics under ambient conditions via depolymerization induced by an externally applied potential and include oxidation and reduction reactions.[66] Primary cleavage of the C-C bonds of waste plastics occurs at the anode, and the process requires a higher anode potential.[67] However, a higher anode potential enhances the depolymerization efficiency at the expense of increased energy consumption. Moreover, a higher current density leads to reduced selectivity and Faraday efficiency. Electrocatalytic reforming of plastics commonly involves alkaline or acidic pretreatment to enhance the solubilization of plastics and facilitate the interaction between catalysts and plastic substrates.[68] However, this pretreatment method has adverse environmental impacts and significantly increases the cost of the reforming process, thereby limiting its large-scale practical application.

### 3.1. Electrocatalytic conversion process and mechanism

Electrocatalytic reactions involve the breaking and formation of chemical bonds and include continuous multistep electron transfer processes.[69] During these reactions, a series of intermediates acquire electrons from the catalyst surface to participate in the reaction (**Figure 10a**). The strength of the interaction between the intermediate and electrocatalyst is characterized by the free energy of adsorption ($\Delta G_{intermediate}$), which is primarily determined by the electronic structure of the electrocatalyst.[70] The adsorption is highly sensitive to the surface properties and the nature of the catalyst–electrolyte interface. Moreover, investigating the relationship between the catalytic activity and the adsorption process of the intermediates provides valuable insights into the design of electrocatalysts (**Figure 10b**). The intermediates are intricately bound, and decoupling the binding among the intermediates is challenging. Notably, the adsorption energy of the hydrogen-containing intermediates exhibited a linear relationship with the adsorption energy of the central atom A ($\Delta G_A$) owing to the similar nature of chemical bonding (**Figure 10c**).[71] This finding has significant implications for the design of catalytic materials. Furthermore, the thermodynamics and kinetics of the adsorption and desorption of key intermediates play a crucial role in achieving highly selective plastic reforming processes.

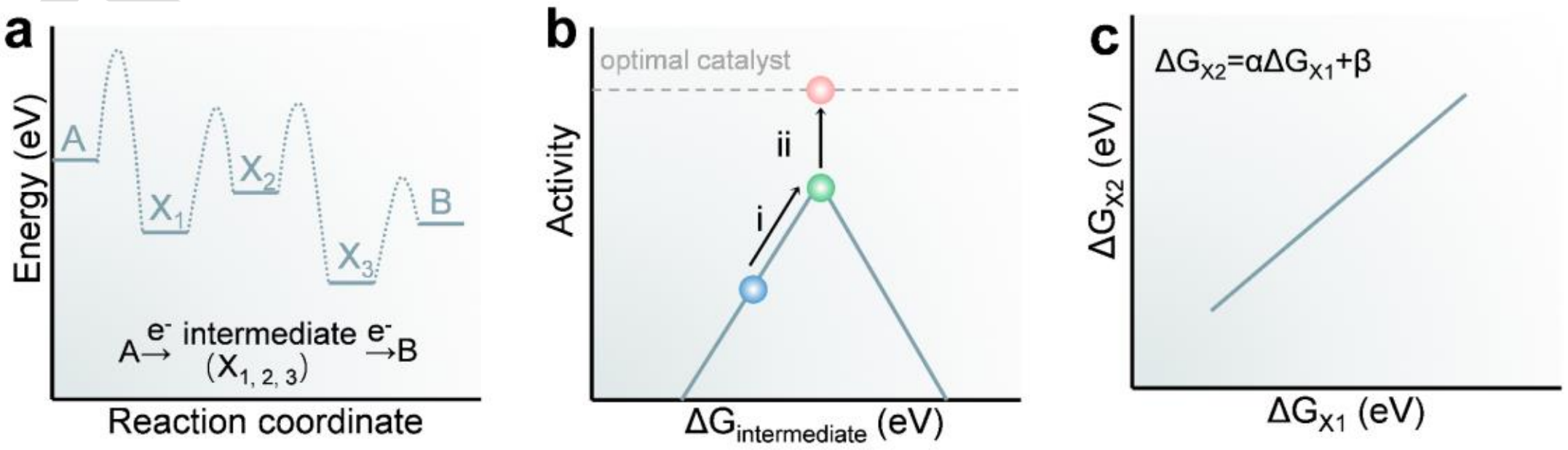

**Figure 10.** Mechanism of plastics electrocatalytic conversion. (a) Free energy diagram showcasing the reaction intermediates and reaction pathways. (b) Volcano-type diagram that establishes a correlation between the adsorption energy of key intermediates and the catalytic activity. (c) Scaling relationship between the adsorption energies of intermediates.

Currently, the electrocatalytic conversion processes for plastic depolymerization predominantly involve heterogeneous reactions. These reactions primarily involve direct integrated reactions, leading to the generation of intermediates or products, as well as local self-feeding tandem reactions involving subsequent reactions with the adsorbed products.[72] The electrocatalytic conversion reaction for plastic depolymerization is a relatively complex process involving the simultaneous activation of multiple molecules.[73] It is imperative for catalysts to be able to independently regulate the adsorption of different intermediates in each component reaction and facilitate their interactions to efficiently form new bonds. To achieve efficient and selective electrocatalytic processes, researchers have developed various strategies that primarily involve the construction of multiple functional sites and the introduction of new degrees of freedom to manipulate intermediate adsorption (**Figure 11**).[74] Such advancements are important for guiding the design of electrocatalytic systems for plastic reforming.

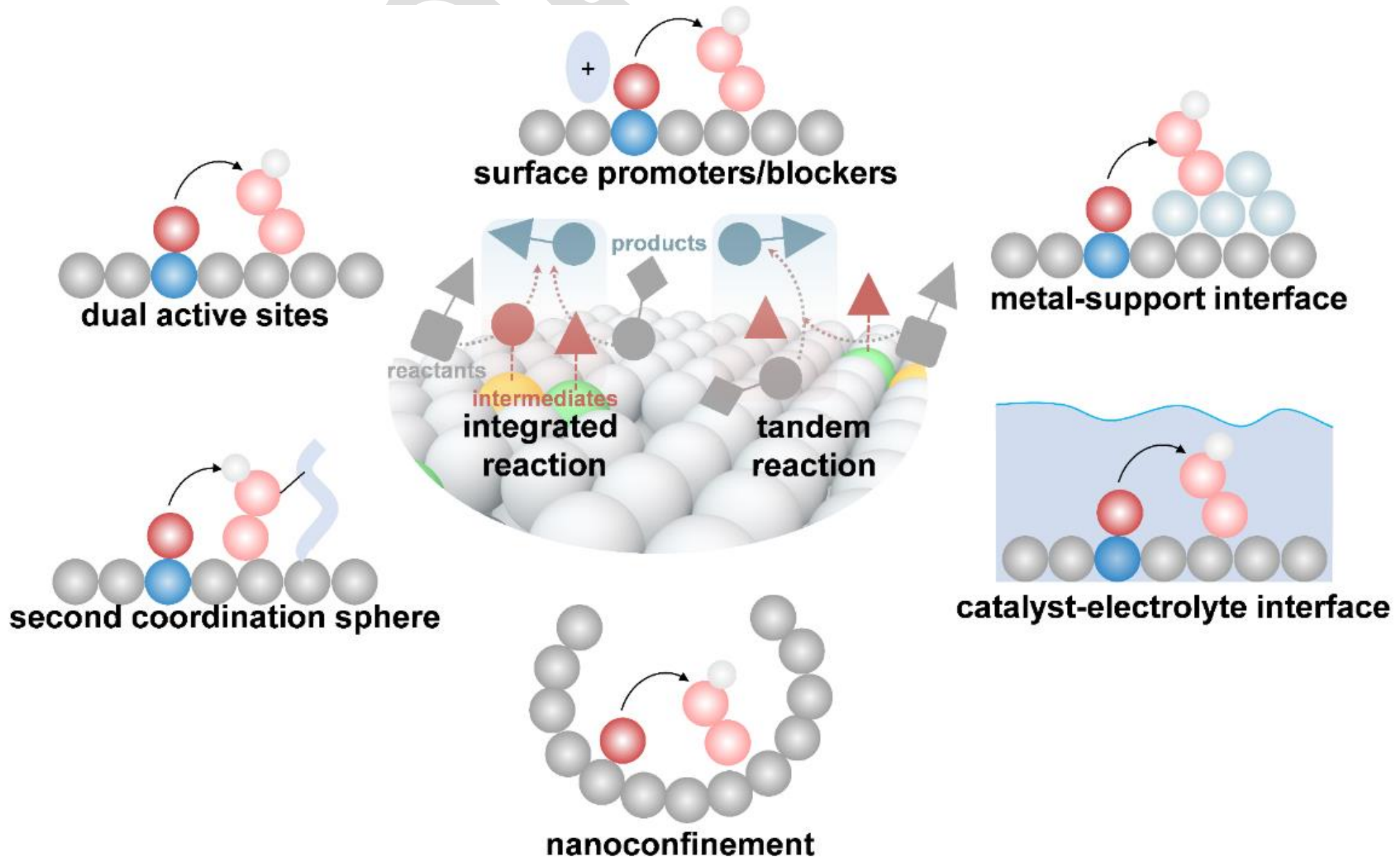

**Figure 11.** Reaction types and catalyst design strategies for electrocatalytic conversion of plastics.

### 3.2. Overview of Recent Advances

*3.2.1. Electrocatalytic conversion into $C_1$ products*

The utilization of higher oxidation potentials allows the rapid depolymerization of plastic polymers, enabling the conversion of plastics into valuable chemicals. However, this approach often leads to lower product selectivity and the formation of numerous by-products, which hinder the efficient separation and utilization of the desired chemicals.[75] To address this challenge, Duan et al.[76] employed a non-noble metal-based cobalt-nickel phosphide electrocatalyst for upcycling waste PET plastics. Technoeconomic feasibility analysis (TEA) was conducted to assess the economic viability of the electrocatalytic PET upcycling strategy. As illustrated in **Figure 12a**, the $CoNi_{0.25}P$ catalyst exhibited partial surface oxidation and the formation of a core-shell structure during the hydrogen evolution reaction. An anion-exchange membrane was used to assemble the membrane electrodes (**Figure 12b**). The introduction of ethylene glycol into the electrolyte revealed that the $CoNi_{0.25}P$/NF catalyst outperformed the other catalysts, exhibiting a lower onset potential and higher current density (**Figure 12c**). At a lower oxidation potential, EG was efficiently oxidized to formate with a Faradaic efficiency of 90% and a formic acid yield of up to 4.1 $mmol\cdot cm^{-2}\cdot h^{-1}$ (**Figure 12d,e**). The Sankey diagram in **Figure 12f** shows that 1 kg of PET plastic produces 389.2 g of formic acid and 818.5 g of PTA. In summary, the electrocatalytic PET reforming strategy involves a multistep process for obtaining KDF (**Figure 12g**). Device-based plastic electrocatalysis enables higher current densities and Faradaic efficiencies at lower oxidation potentials. However, it is important to note that the use of electrocatalysis to convert high-value high-carbon fuels into low-value $C_1$ products may result in the waste of energy.

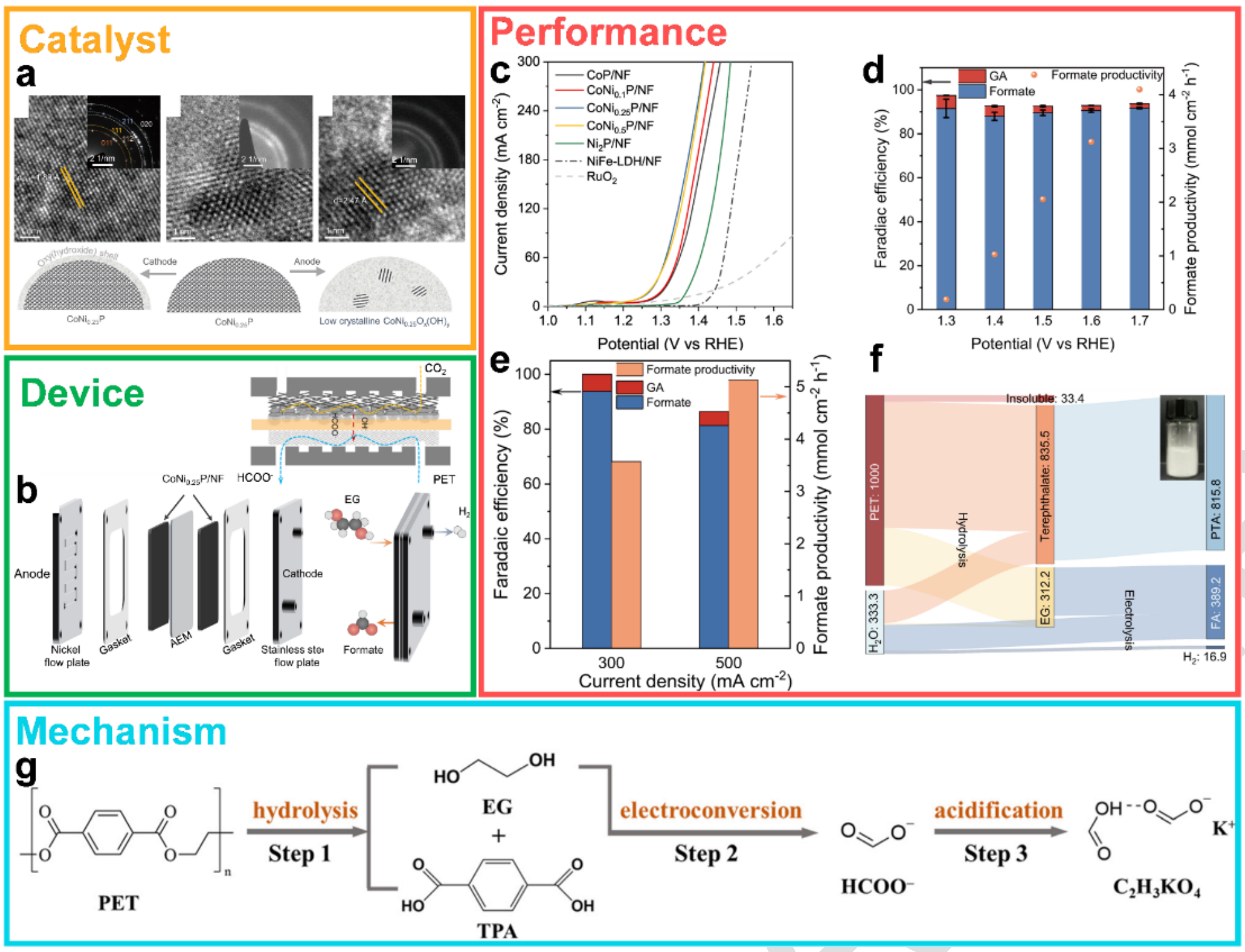

**Figure 12.** (a) TEM images and corresponding SAED patterns of $CoNi_{0.25}P$, and schematic illustration of the structural evolution of $CoNi_{0.25}P$ catalyst under reaction conditions. (b) The MEA setup for paired HER(−)//EG oxidation(+). (c) LSV curves (85% iR corrected) for EG oxidation. (d) FE and productivity of formate on $CoNi_{0.25}P$/NF at different potentials. (e) Faradaic efficiency and productivity as function of current density for EG oxidation. (f) Sankey diagram for the mass flow of PET upcycling. Photograph of separated high-purity PTA is shown (inset). (g) Scheme for electrocatalytic PET conversion into TPA and KDF via the three-step pathway. Reproduced with permission.[76] Copyright 2021, Springer Nature.

### *3.2.2. Electrocatalytic conversion into $C_{2+}$ products*

Electrochemical oxidation is a sustainable and promising approach for the direct conversion of waste plastics into valuable fine chemicals.[77] However, the current products obtained are predominantly low-value $C_1$ chemicals, which is not in line with the principles of a circular economy.[78] To achieve the highly selective production of higher-value $C_{2+}$ products, Duan et al.[79] developed an Au/$Ni(OH)_2$ synergistic catalyst with dual active sites for enhanced adsorption and catalysis. Their strategy facilitated the mass transfer, adsorption, and catalytic processes of the reactants on the catalyst surface, enabling the selective conversion of biomass resources (glycerol) and waste plastics (PET) into high-value lactic acid and glycolic acid at an industrial-scale high current density (>300 mA·cm$^{-2}$) with a designed diaphragm-free stack (**Figure 13a-d**). Notably, at high conversion rates, the lactic acid and glycolic acid yields were 75% and 91%, respectively, in the presence of Au/Ni(OH)$_2$ (**Figure 13e**). In situ infrared spectroscopy was conducted to elucidate the reaction pathway for glycerol oxidation (**Figure 13f**). Glycerol was oxidized to GLD/DHA, leading to spontaneous lactic acid formation, while ethylene glycol was converted to glycolic acid via oxidation. (**Figure 13g**). This study offers a novel approach for the valorization of carbon resources such as plastics and biomass into high-value C$_{2+}$ products. However, high oxidation potentials also lead to higher energy consumption. Thus, achieving fast and precise cleavage of C-C bonds at a low current remains significantly challenging.

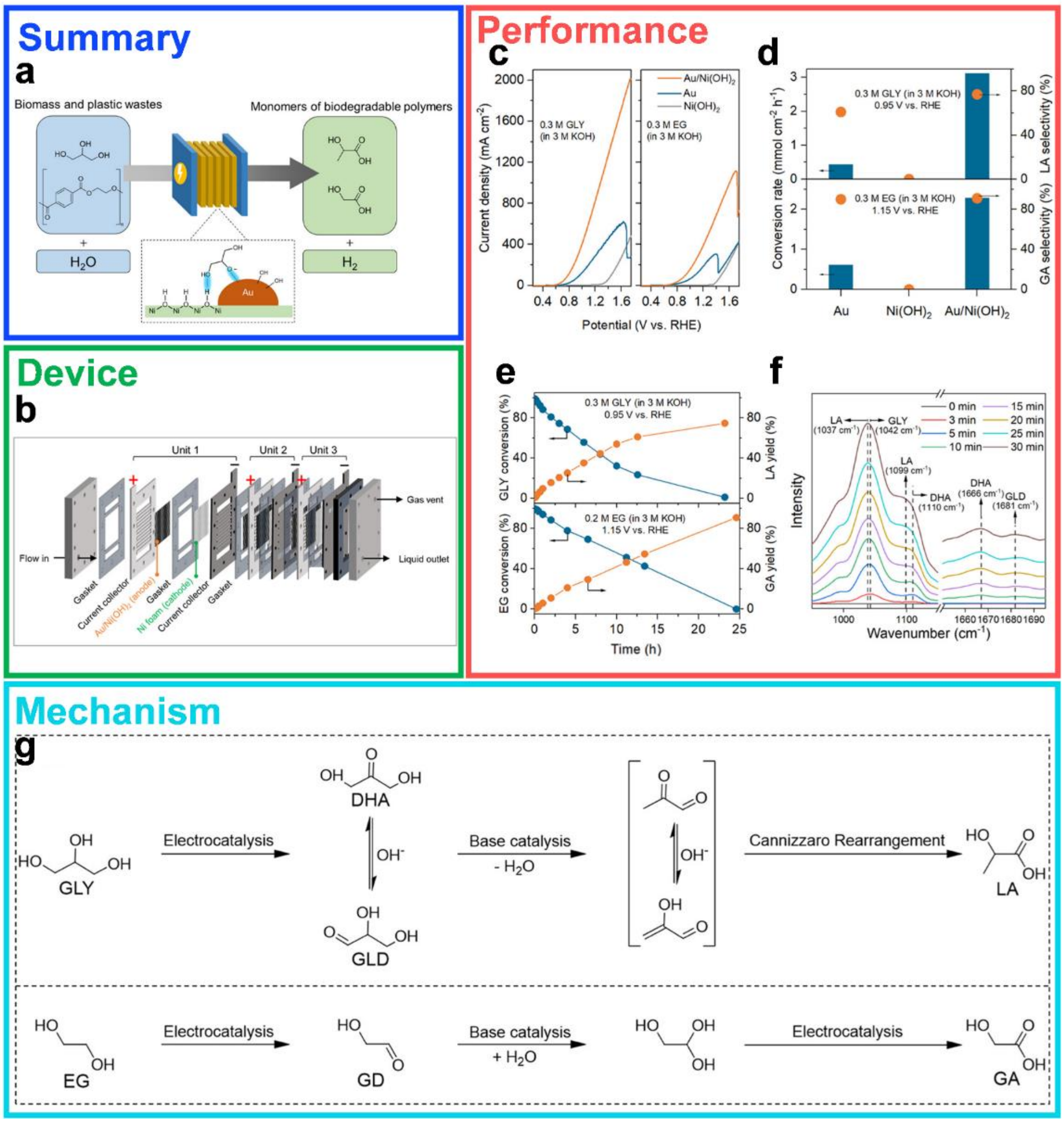

**Figure 13.** (a) Electrocatalytic upcycling of biomass and plastic wastes to monomers of biodegradable polymers and H$_2$. (b) Photograph and schematic illustration of the stacked membrane-free flow electrolyzer with three units. (c) LSV curves of GLY (left) and EG (right)

electrooxidations at a scan rate of 10 mV/s in 3 M KOH with 0.3 M substrates over different catalysts. (d) Conversion rate of GLY (top) and EG (down), and the corresponding selectivity of LA (top) and GA (down) over different catalysts. (e) Kinetic curves for GLY (top) and EG (down) electrooxidations over Au/Ni(OH)$_2$. (f) Infrared spectra recorded during glycerol oxidation on Au/Ni(OH)$_2$. (g) Proposed reaction mechanisms for electrooxidation of GLY-to-LA (top) and EG-to-GA (down). Reproduced with permission.[79] Copyright 2023, American Chemical Society.

The electrocatalytic conversion of ethylene glycol, which is derived from PET, into high-value $C_{2+}$ chemicals and hydrogen offers a promising solution to the pressing issue of plastic pollution. Chen et al.[80] studied the electrooxidation of EG to glycolic acid using a Pd-Ni(OH)$_2$ catalyst (**Figure 14a**), demonstrating exceptional performance even at industrial-grade current densities (600 mA·cm$^{-2}$, 1.15 V *vs.* RHE) with a custom-made two-electrode membrane electrode assembly flow electrolyzer (**Figure 14b**). The catalyst exhibited high Faradaic efficiency and selectivity (>85%), and the electrolysis was stable for over 200 h. Pd-Ni(OH)$_2$ exhibited a significantly lower oxidation potential and enhanced current density for ethylene glycol oxidation compared to the oxygen evolution reaction (OER) in alkaline electrolytes (**Figure 14c**). The addition of ethylene glycol substantially decreased the required reaction potential, indicating a more favorable kinetic profile than that of the OER (**Figure 14d**). Pd-Ni(OH)$_2$ showed excellent electrocatalytic performance with a high Faradaic efficiency of 90% and ethylene glycol conversion of 93.2%, with 91.6% selectivity toward glycolic acid (**Figure 14e-g**). The generation of *OH active species was the rate-determining step in this process, with Pd-Ni(OH)$_2$ facilitating the conversion of the intermediates into glycolic acid at a lower potential (**Figure 14h**). This study achieved the highly selective conversion of the PET plastic derivative, ethylene glycol, to glycolic acid at a lower potential, implying a lower application cost. The production of higher amounts of $C_{2+}$ products will be of great significance for the carbon cycle. However, the current electrocatalytic conversion technology is limited to the conversion of monomers or derivatives of plastic products, rendering it difficult to achieve the direct reforming of plastic wastes. Therefore, it is worth considering whether combining electrocatalytic processes with other catalytic reactions can expand the application range of plastic reforming reactions.

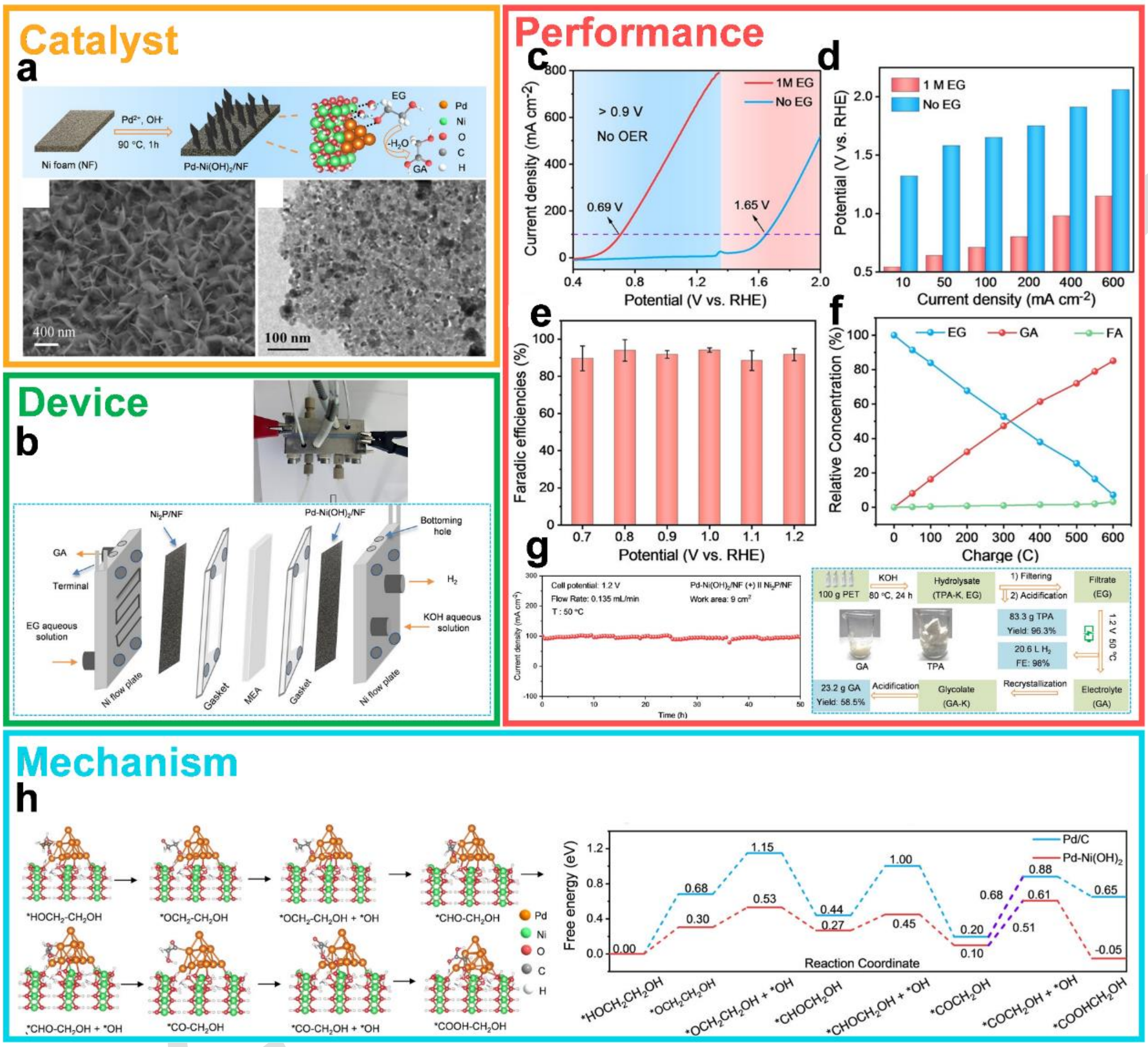

**Figure 14.** (a) Schematic illustration of the synthesis of Pd−$Ni(OH)_2$ on NF, and corresponding SEM images and TEM images. (b) Schematic illustration and photograph of a two-electrode MEA flow electrolyzer for PET electro-reforming. (c) LSV curves and (d) comparisons of the potentials needed to achieve designated current densities for Pd−$Ni(OH)_2$/NF in 1.0 M KOH with and without EG. (e) FE of Pd−$Ni(OH)_2$/NF for GA production at designated voltages. (f) The concentration of EG, FA, and GA during electrolysis. (g) CP curves of Pd−$Ni(OH)_2$/NF(+) and $Ni_2P$/NF(−) for PET electro-reforming, and schematic illustration of the process of PET electro-reforming and product separation. Reproduced with permission.[80] Copyright 2023, John Wiley and Sons.

### *3.2.3 Electrocatalytically coupled multiple reactions for conversion to $C_1$ products*

Coupling waste polymer plastics with other electrocatalytic processes for high-value chemical production is a more efficient approach for plastic upcycling and significantly enhances the production efficiency of valuable chemicals. Ma et al.[81] employed cost-effective and readily available nickel-cobalt-based and tin-based oxide materials for the electrocatalytic oxidation of PET, coupled with the electrocatalytic reduction of $CO_2$, to achieve the selective coproduction of high-value formic acid under lower voltage conditions (**Figure 15a**). $NiCo_2O_4$ nanowires exhibited catalytic activity and selectivity for the electrocatalytic oxidation of PET hydrolyzate, showing a 200-mV reduction in the potential of the oxidation reaction compared to that of water oxidation at a catalytic current density of 50 $mA{\cdot}cm^{-2}$ (**Figure 15b,c**). Moreover, the catalytic reaction exhibited remarkable selectivity, with a Faradaic efficiency of the formic acid product reaching approximately 90% (**Figure 15d**). A two-electrode electrolyzer with $NiCo_2O_4$/CFP and $SnO_2$/CC electrodes showed a decrease in the driving voltage upon the addition of 0.1 M PET hydrolysate, indicating effective reduction in the energy consumption associated with the OER (**Figure 15e**). The formic acid Faradaic efficiency reached approximately 85%, remaining stable at battery voltages below 2.0 V and thus highlighting the favorable reaction stability (**Figure 15f**). Investigation of the reaction mechanism revealed that oxalic acid was the primary product of glycolic acid oxidation, while formic acid was the main product of ethylene glycol oxidation over the $NiCo_2O_4$ electrocatalyst, suggesting that glycolic acid was not a significant intermediate in ethylene glycol oxidation (**Figure 15g**). A proposed reaction pathway for the oxidation of ethylene glycol over $NiCo_2O_4$ electrocatalysts is presented based on these findings (**Figure 15h**). The coupling of waste plastic electrocatalysis with electrocatalytic $CO_2$ conversion would enable the efficient upcycling of PET waste plastics, enhance the production rate of formic acid, and reduce $CO_2$ emissions, thereby fostering the development of a carbon circular economy. It should be noted that although coupling with $CO_2$ electrocatalysis can expand the application areas of PET plastic upcycling to a certain extent, the conversion to low-carbon chemical products will still lead to wastage of resources and limit the efficiency.

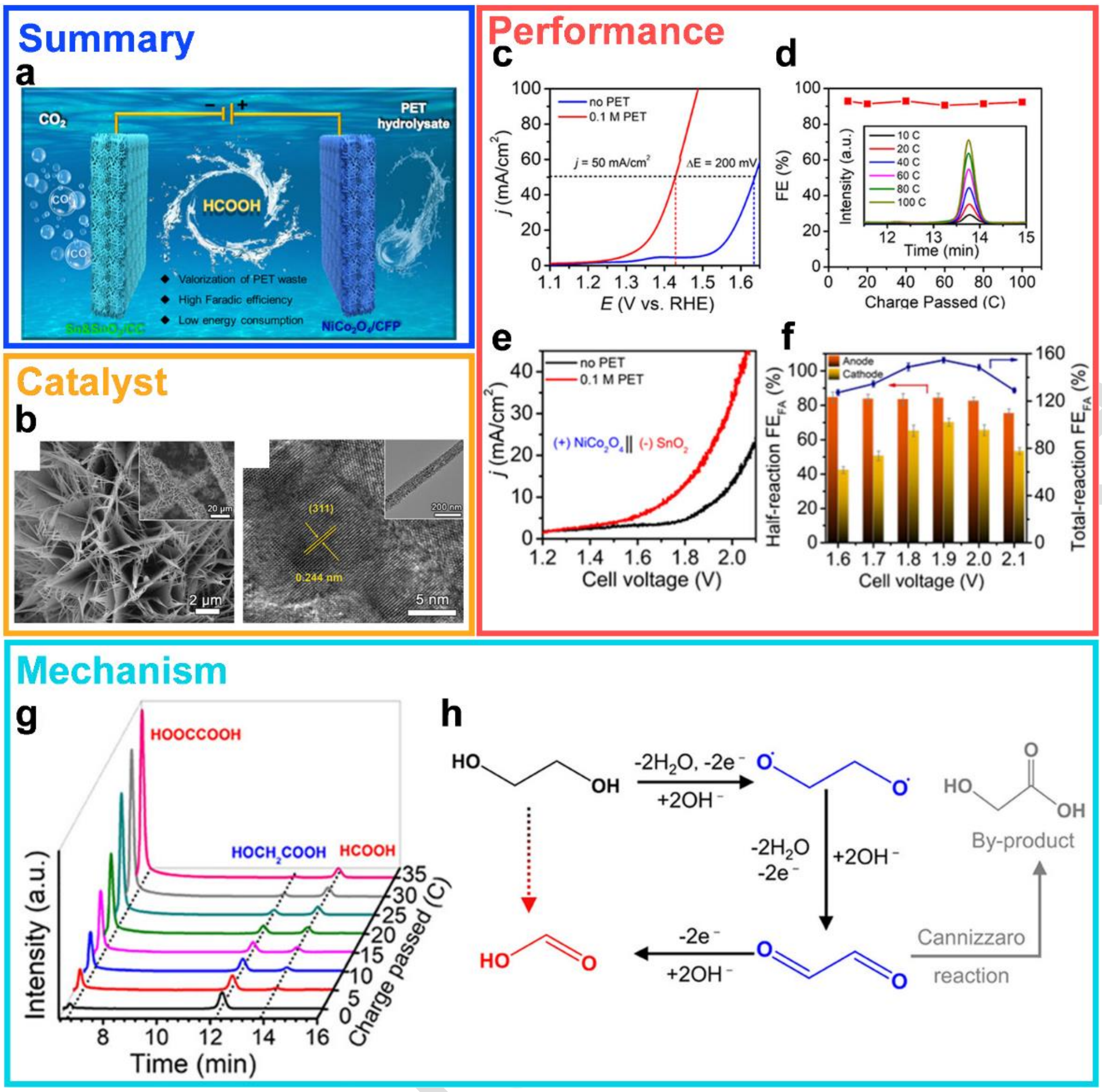

**Figure 15.** (a) Concurrent electrochemical conversion of anodic pet hydrolysate and cathodic $CO_2$ to formic acid. (b) SEM and TEM images of the as-prepared $NiCo_2O_4$ material. (c) LSV curves at a scan rate of 10 mV/s of $NiCo_2O_4$/CFP in 1 M NaOH solutions with and without 0.1 M PET hydrolysate. (d) FEs of PET hydrolysate oxidation to formic acid with different charges passed at the applied potential of 1.45 V; the inset shows the corresponding high-performance liquid chromatography (HPLC) chromatograms. (e) LSV curves for the $SnO_2$||$NiCo_2O_4$ electrolytic cell with and without the existence of 0.1 M PET hydrolysate in the anode cell. (f) Faradaic efficiencies of $NiCo_2O_4$/CFP for PET hydrolysate oxidation and $SnO_2$/CC for the $CO_2$RR to produce formic acid at different applied cell voltages. (g) HPLC of the glycolic acid oxidation products at a constant potential of 1.45 V with different amounts of charge passed. (h) Proposed reaction pathway for the electrochemical oxidation of ethylene glycol to formic acid. Reproduced with permission.[81] Copyright 2022, American Chemical Society.

### *3.2.4 Electrocatalytically coupled multiple reactions for conversion to into $C_{2+}$ products*

Converting carbon- and hydrogen-rich waste plastics into higher-value $C_{2+}$ products can avoid energy wastage to a certain extent and enable high-value utilization of carbon resources, which is of great significance for realizing the upcycling process of waste plastics. Park et al.[82] reported a solar-driven biocatalytic system using non-recyclable real-world PET microplastics as electronic feedstocks to achieve C–H oxygen functionalization of bonds, amination of reducing C=O bonds, and trans-hydrogenation of C=C bonds (**Figure 16a**). Worm-like α-$Fe_2O_3$ structures were synthesized via hydrothermal deposition, and the influence of Zr doping on plastic reforming was investigated (**Figure 16b**). Results showed that the generation of formate and acetate was faster using a Zr:α-$Fe_2O_3$ photoanode than using a α-$Fe_2O_3$ photoanode in the voltage range of 0.8–1.2 V *vs.* RHE (**Figure 16c**). Moreover, at the same PET concentration, the Zr:α-$Fe_2O_3$ photoanode exhibited higher photoanode currents than α-$Fe_2O_3$ (**Figure 16d**). Consequently, this approach facilitated a stable biosynthetic reaction with a $TTN_{rAaeUPO}$ of 271,000 (7 h) (**Figure 16e**). This study proposes an electrocatalytically coupled bioenzymatic reaction that combines environmental remediation and biocatalytic photosynthesis for the sustainable synthesis of high-value chemicals. The multifaceted design based on an electrocatalytic process significantly enhances the efficiency of the selective conversion of waste plastics into high-carbon products. Furthermore, the multifunctional characteristics of the approach broaden its potential for practical applications. The current electrocatalytic processes mainly focus on the catalytic activity of PET because of its alkaline hydrolysis characteristics. However, research on the electrocatalytic effects of other high-polymer plastics is still lacking, which limits the application range of electrocatalytic reforming. Thus, exploring the electrocatalytic potential of other high-polymer plastics is essential for future development.

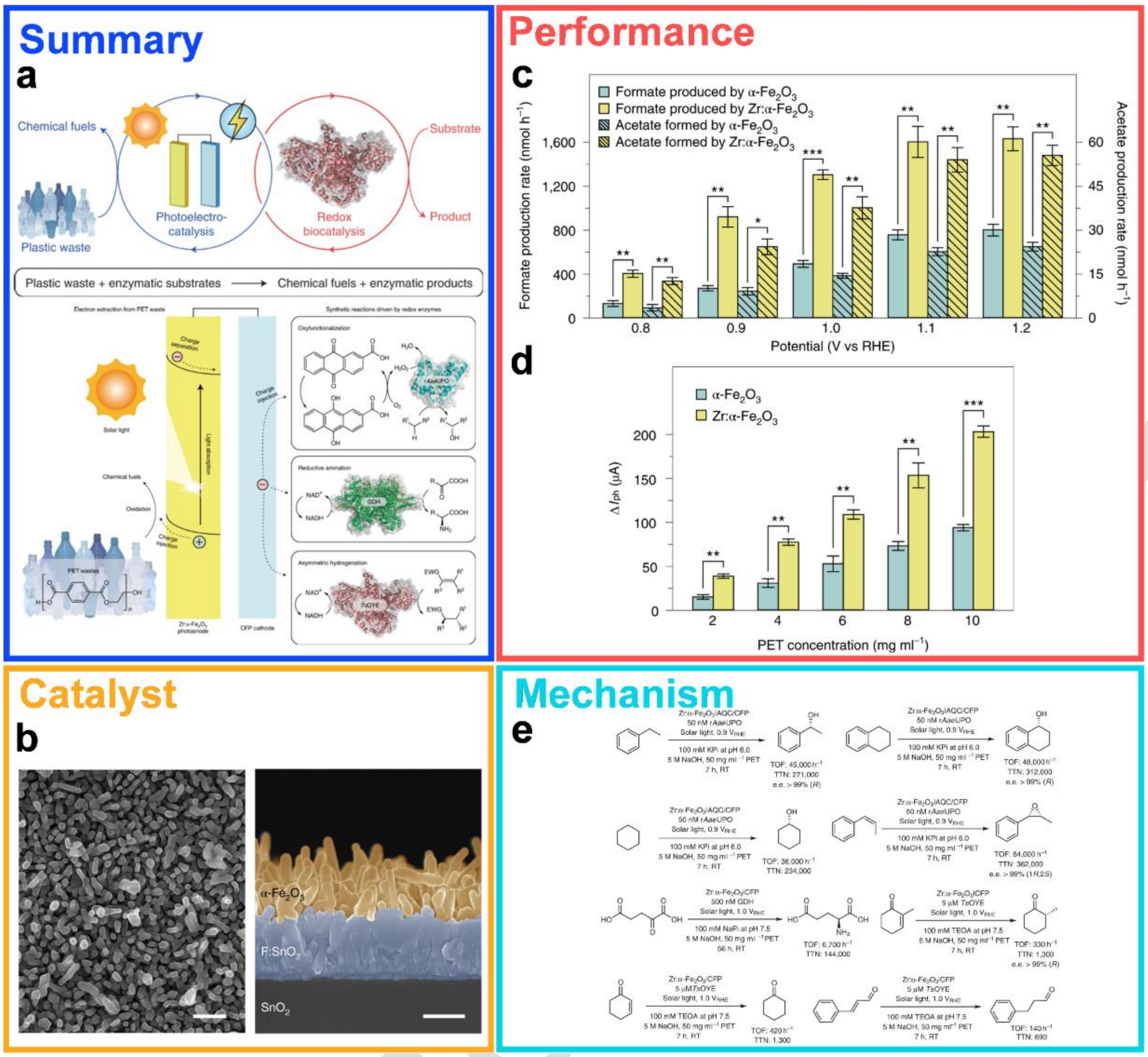

**Figure 16.** (a) Schematic diagram of solar-powered photoelectrochemical biosynthetic reactions using non-recyclable real-world PET microplastics. (b) Plan-view and cross-sectional-view SEM images of α-$Fe_2O_3$ electrode. (c) Potential-dependent production rates of formate and acetate driven by α-$Fe_2O_3$ and Zr:α-$Fe_2O_3$ photoanodes. (d) Comparison of $\Delta I_{ph}$ of the haematite-based photoanodes at 1.1 $V_{RHE}$. (e) Substrate scope of BPEC reactions using real-world PET microplastics. Reproduced with permission.[82] Copyright 2022, Springer Nature.

*3.2.5 Other notable electrocatalytic plastic reforming technologies*

Electrocatalytic waste plastic upcycling is an appealing strategy, and current research focuses on achieving high current densities at low anodic potentials while maintaining a high Faradaic efficiency. A recent study by Chen et al.[83] introduced a novel electrocatalytic strategy using a Pd-modified Ni foam catalyst to directly convert waste PET into carbonate. The catalyst demonstrated excellent electrocatalytic activity for carbonate production, achieving a current density of 400 mA·cm$^{-2}$ at 0.7 V *vs.* RHE. Furthermore, the system exhibited high selectivity (95%) and Faradaic efficiency (93%). Notably, this strategy eliminates the need for alkali treatment of plastics and exhibits favorable economic and environmental characteristics. Considering the current research progress in plastic electrocatalysis (**Table 4**), it is evident that achieving highly selective conversion of waste into high-carbon, high-value-added chemicals is a future development direction in this field.

**Table 4.** Electrocatalytic conversion of plastics over different catalysts.

| Category | Catalysts | Plastics | Potential (V vs. RHE) | FE (%) | Products | Selectivity | Refs. |
|---|---|---|---|---|---|---|---|
| $C_{2+}$ | $CoNi_{0.25}P$ | PET | 1.70 | 91.3 | potassium diformate and terephthalic acid | >80% | [76] |
| | $Au/Ni(OH)_2$ | PET | 1.15 | 96 | lactic acid and glycolic acid | 77 and 91% | [79] |
| | $Pd-Ni(OH)_2/NF$ | PET | 1.15 | 86 | glycolic acid | 91.6 % | [80] |
| | Pd/Ni | PET | 0.70 | 93 | carbonate | 95% | [83] |
| $C_1$ | $NiCo_2O_4$ | PET | 1.55 | 90 | formic acid | >90% | [81] |
| | CuO | PET | 1.50 | 90 | formic acid | 90% | [84] |
| | Ni/CoP | PET | 1.30 | 96 | formic acid | 96% | [85] |

## 4. Photocatalytic conversion of plastics

The photoreforming of plastics utilizing solar energy is a promising technology for processing plastic wastes into higher-value chemicals, leveraging renewable solar energy.[86] This approach offers distinct efficiency and economic advantages over plastic photodegradation and closely aligns with the principles of sustainable development (**Table 5**). Compared to thermal catalytic conversion and electrocatalytic conversion, the photoreforming of plastics does not require additional energy input, resulting in lower costs and ensuring greater economic feasibility.[87] Currently, plastic photoreforming focuses on the production of high-value fuels, materials, and chemicals.[88] Achieving selective generation of these high-value products is the primary research objective in the field of plastic photoreforming.

**Table 5.** Electrocatalytic conversion of plastics over different catalysts.

| Category | Catalysts | Plastics | Light sources | Times | Products | Refs. |
|---|---|---|---|---|---|---|
| | $TiO_2$/CuPc | PS | Fluorescent light | 200 h | $CO_2$ | [89] |
| Photodegradation | Vitamin C-$TiO_2$ | PVC | UV light | 216 h | $CO_2$ | [90] |
| | $TiO_2$ | PE | Solar light | 200 h | $CO_2$ | [91] |
| | p-Toluenesulfonic | PS | Ultraviolet LEDs | 24 h | formic acid | [92] |
| Photoreforming | $UO_2(NO_3)_2 \cdot 6H_2O$ | PS | 460 nm light | 72 h | benzoic acid | [93] |
| | ZnO/$UiO_{66}$-$NH_2$ | PVC/PLA | Solar light | 35 h | acetic acid | [94] |

### 4.1 Photocatalytic conversion process and mechanism

Photocatalytic reactions consist of three main processes. First, light irradiation occurs, upon which photocatalysts absorb photons and produce photogenerated electron-hole pairs. Second, the photogenerated carriers separate and migrate. Finally, a redox reaction of the photogenerated carriers occurs on the catalyst surface (**Figure 17a**).[95] From a thermodynamic perspective, the energies of the photogenerated carriers must align with the redox potential of the desired product, necessitating a suitable energy band structure (**Figure 17c**).[96] However, the absorption process requires photons with energies higher than the bandgap of the catalyst.[97] Thus, the bandgap of the photocatalyst should be as narrow as possible to maximize the absorption of visible light, which constitutes 43% of the solar spectrum (**Figure 17b**).[98] In plastic photoreforming, a lower potential is required for plastic oxidation, allowing for a wider range of catalyst choices. Therefore, the selective photoreforming of plastics is primarily associated with catalyst design.

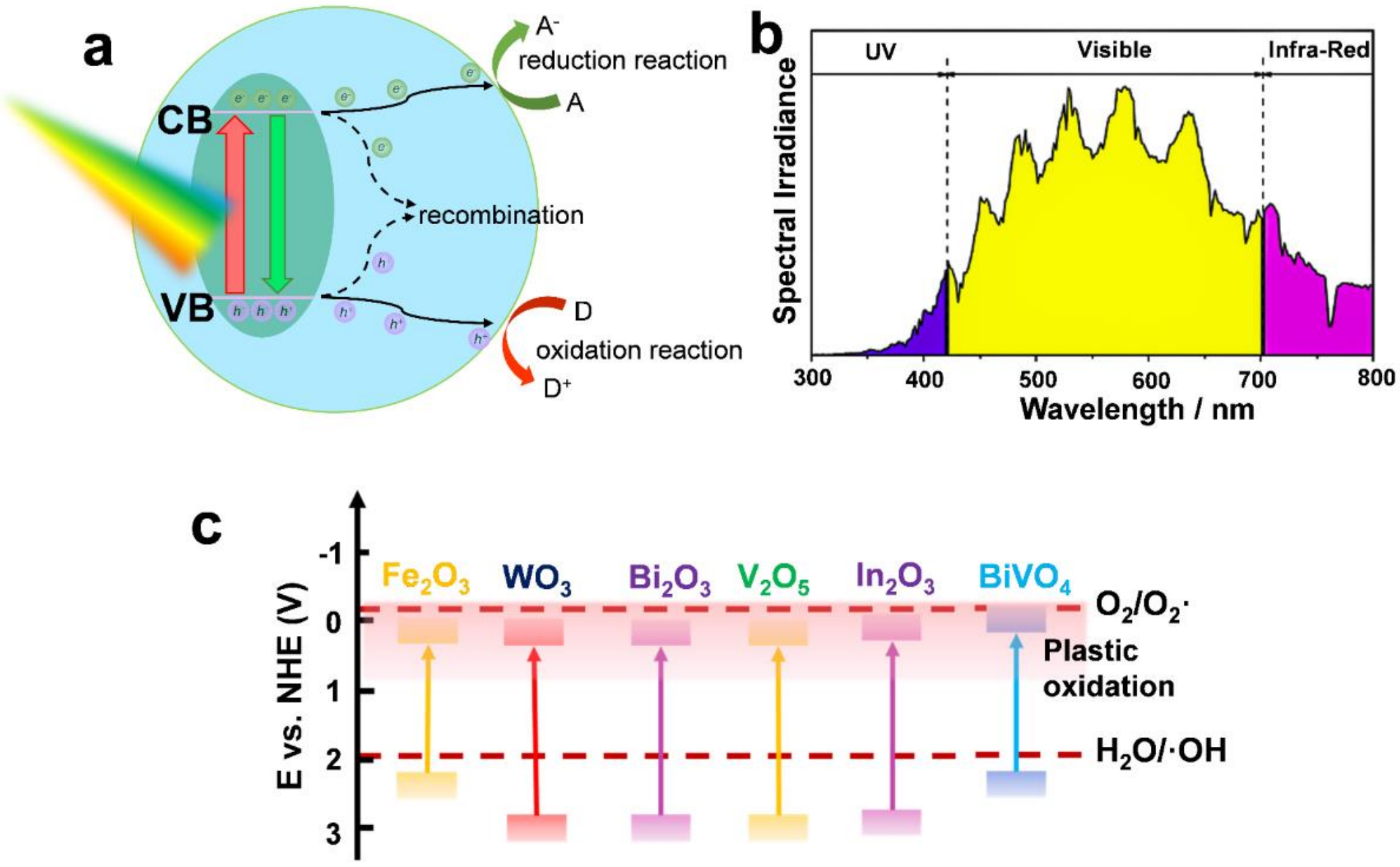

**Figure 17.** (a) Schematic diagram of photocatalytic processes. (b) Solar spectrum energy distribution. (c) Bandgaps and band edge positions of various semiconductors on a potential scale versus the normal hydrogen electrode (NHE).

To achieve the highly selective photoreforming of plastics, it is necessary to explore catalyst optimization strategies with broader applicability. These strategies primarily involve the following. 1) Defect engineering: This approach focuses on introducing defects or heteroatoms to optimize the catalyst performance. By constructing active sites on the catalyst surface and optimizing its energy band structure, the energy barrier can be significantly reduced and the catalytic activity at the reaction center during the adsorption process can be improved (**Figure 18a**).[99] 2) Loading of co-catalysts: Loading metals as co-catalysts can enhance charge transfer at the interface. Additionally, the surface plasmon resonance effect can be harnessed to adjust the redox potential of the photogenerated carriers (**Figure 18b**).[100] 3) Material hybrid strategies: Hybridizing photocatalysts with nonmetallic components, such as carbon-based materials or metal oxides, can leverage the unique properties of different materials. This provides researchers with a powerful tool for manipulating light harvesting strategies, adsorption capacity, catalytic activity, and selectivity (**Figure 18c**).[101] 4) Band engineering regulation: By constructing a properly matched energy band structure, an electric field can be created on the surface or interface, promoting the transfer of photogenerated carriers and significantly enhancing the catalytic activity (**Figure 18d**).[102]

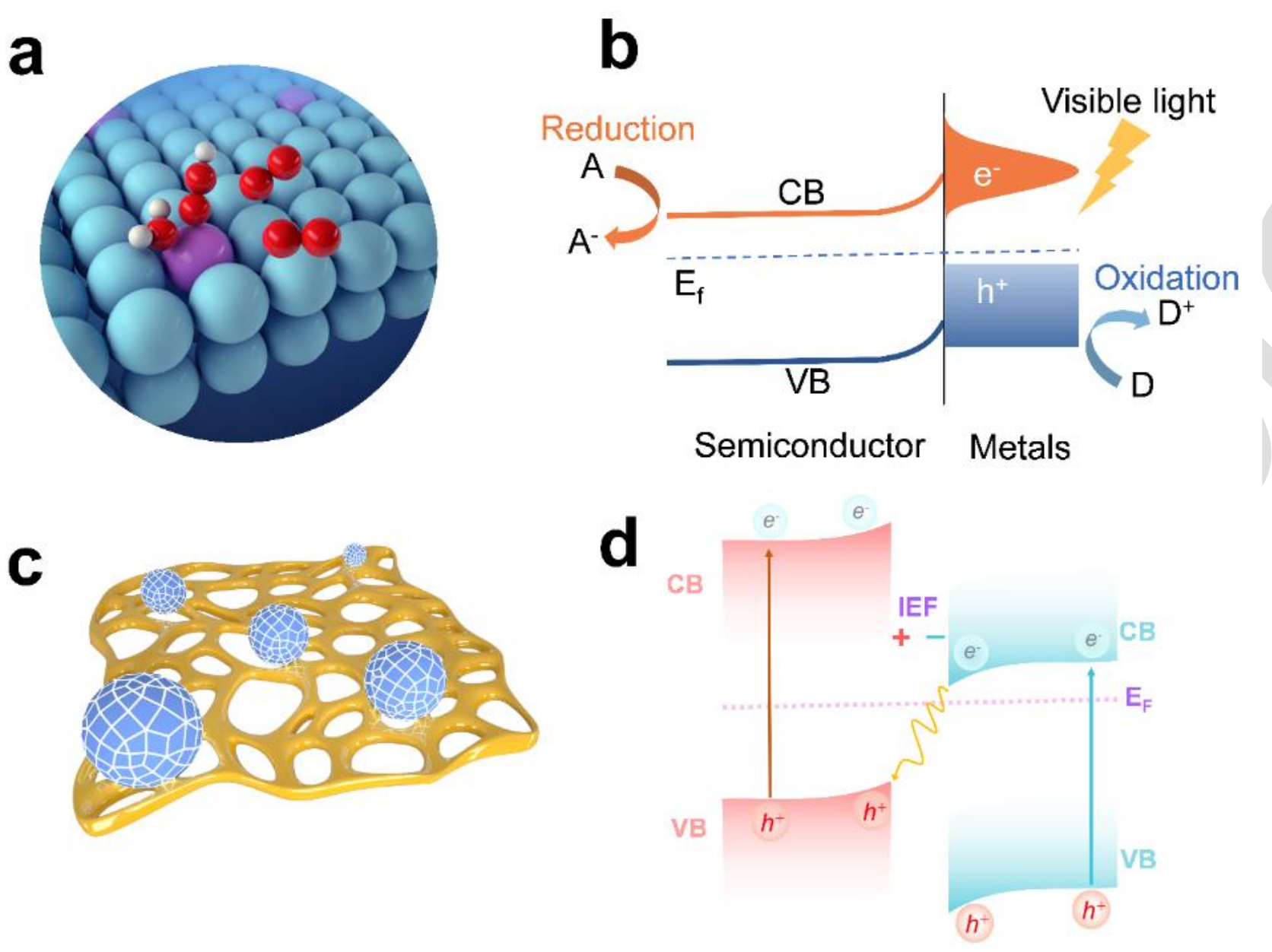

**Figure 18.** Schematic diagram of defect engineering (a), surface plasmon resonance effect (b), materials hybrid strategies (c) and constructing heterojunction (d).

## 4.2 Overview of Recent Advances

### *4.2.1 Photocatalytic conversion into fuels*

Plastic photoreforming offers a promising approach for generating green fuels, such as $H_2$ and CO.[103] One of the key advantages is the utilization of waste plastics as sacrificial agents for hole scavenging, which effectively suppresses the recombination of photogenerated charges and enhances the overall photocatalytic performance.[104] Furthermore, from a thermodynamic standpoint, the preparation of green fuels via plastic photoreforming requires less energy than water oxidation.[105] This significant thermodynamic advantage stems from the lower energy requirements for the oxidation of waste plastics. Consequently, the production of fuels via plastic photoreforming has garnered considerable attention. In a notable study, Reisner et al.[106] reported the use of a novel CdS/$CdO_x$ quantum dot catalyst (**Figure 19a**) for the direct phototransformation of plastics in an alkaline solution to produce hydrogen. This breakthrough work demonstrates the successful conversion of various real-world plastics, including PLA, PET, and PUR, into organic products (such as formate, acetate, and pyruvate) through hole oxidation, where photogenerated electrons reduce protons to generate hydrogen (**Figure 19b**). The results revealed that the prepared CdS-based catalyst exhibited hydrogen production activities of 64.3, 3.42, and 0.85 $mmol_{H2}\cdot g_{CdS}^{-1}\cdot h^{-1}$ for PLA, PET, and PUR, respectively (**Figure 19c**). Moreover, continuous and stable hydrogen production from the photoreforming of real PET water bottles was achieved with an activity of 4.13 $mmol_{H2}\cdot g_{CdS}^{-1}\cdot h^{-1}$ (**Figure 19d**). Simultaneous hydrogen production from plastic photoreforming not only represents a viable approach for converting environmental wastes into clean energy but also holds promise for addressing sustainability challenges. Although this process exhibits excellent hydrogen production performance, the complexity of the organic products limits their separation and utilization, which is against the ultimate goal of realizing the carbon cycle.

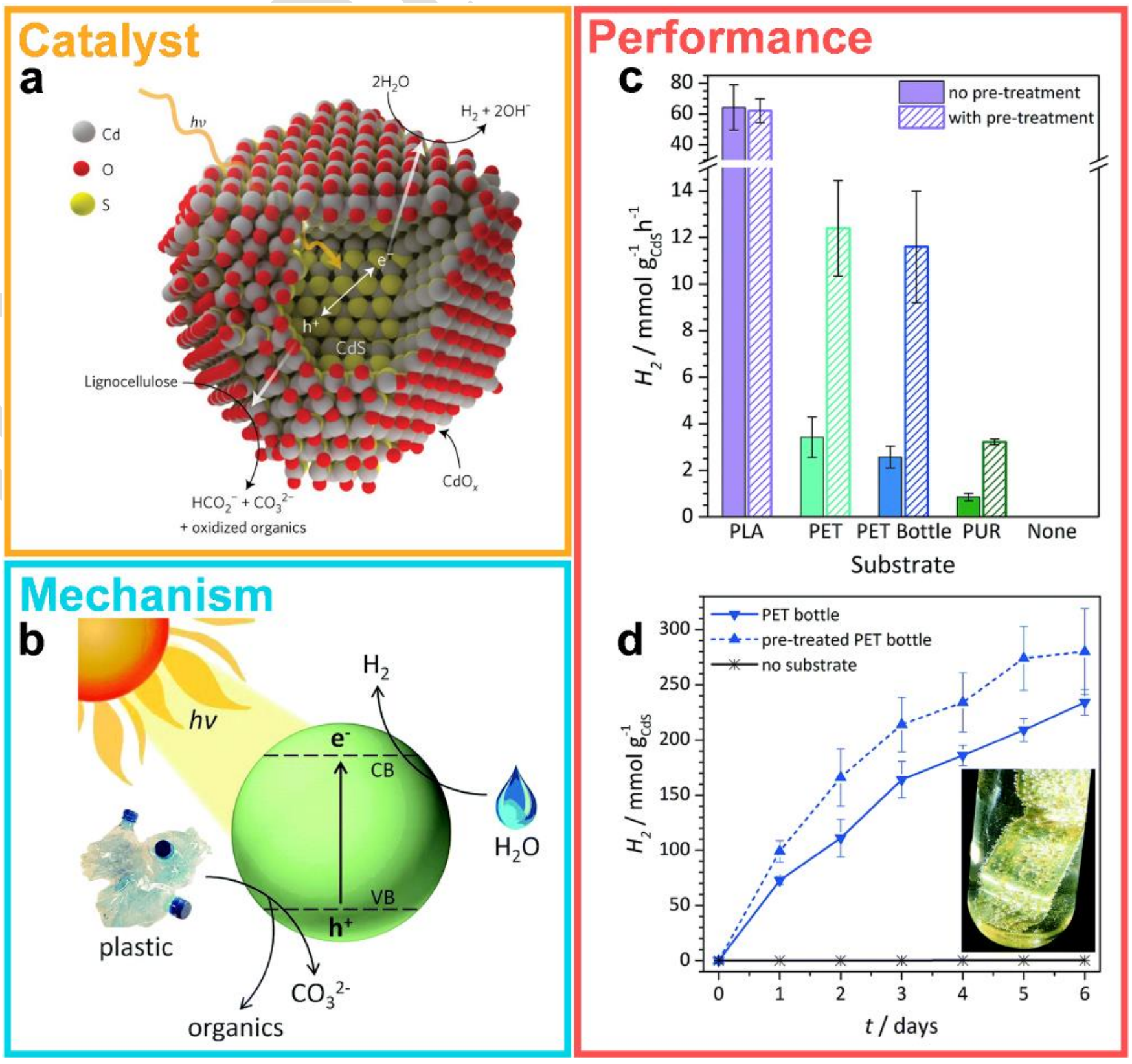

**Figure 19.** (a) The photocatalyst consisted of a semiconductor CdS coated with $CdO_x$. Light absorption by CdS generates electrons and holes, which travel to the $CdO_x$ surface and mediate proton reduction and plastic oxidation, respectively. Reproduced with permission.[107] Copyright 2017, Springer Nature. (b) Diagram of the plastic photosynthesis process with a CdS/$CdO_x$ quantum dot photocatalyst in alkaline aqueous solution. (c) Photosynthesis from plastics into $H_2$ using CdS/$CdO_x$ QDs under simulated solar light. (d) Long-term photosynthesis from PET bottles to $H_2$ using CdS/$CdO_x$ QDs under simulated sunlight. Inset: photograph of PET bottle sample, $H_2$ bubbles are visible on the plastic surface. Reproduced with permission.[106] Copyright 2018, The Royal Society of Chemistry.

In addition to green hydrogen production, the conversion of plastics into CO fuels aligns well with the carbon circular economy strategies. Recently, Sun et al.[108] introduced a novel plastic photoreforming technology that utilized solar energy to achieve the sustainable conversion of plastic into syngas under ambient conditions (**Figure 20a**). By employing ultrathin Co-$Ga_2O_3$ nanosheets, the upcycling of non-recyclable plastics was successfully achieved under mild conditions (**Figure 20b**). The results demonstrate the effective photoreforming of commercial plastics, such as PE plastic bags, PP plastic boxes, and PET plastic bottles, into CO, while simultaneously photoreducing $H_2O$ to $H_2$. Notably, PE plastic bags can be efficiently transformed into syngas using Co-$Ga_2O_3$ nanosheets in the presence of $H_2O$, with $H_2$ and CO production rates reaching 647.8 and 158.3 $\mu mol \cdot g^{-1} \cdot h^{-1}$, respectively (**Figure 20c-e**). The photochemical conversion of commercial plastics and water into syngas involves three key processes: photon absorption, $H_2O$ splitting to produce $H_2$ and $O_2$, and photodegradation of plastics into $CO_2$ followed by CO formation through the *COOH intermediate (**Figure 20f**). This method offers a promising approach for transforming waste plastics into green fuels, contributing to the mitigation of white pollution and energy crises. However, compared to high-carbon chemicals, the added value of CO is low. Therefore, developing efficient methods for converting plastics into high-value-added products requires more attention for further advancements in this field.

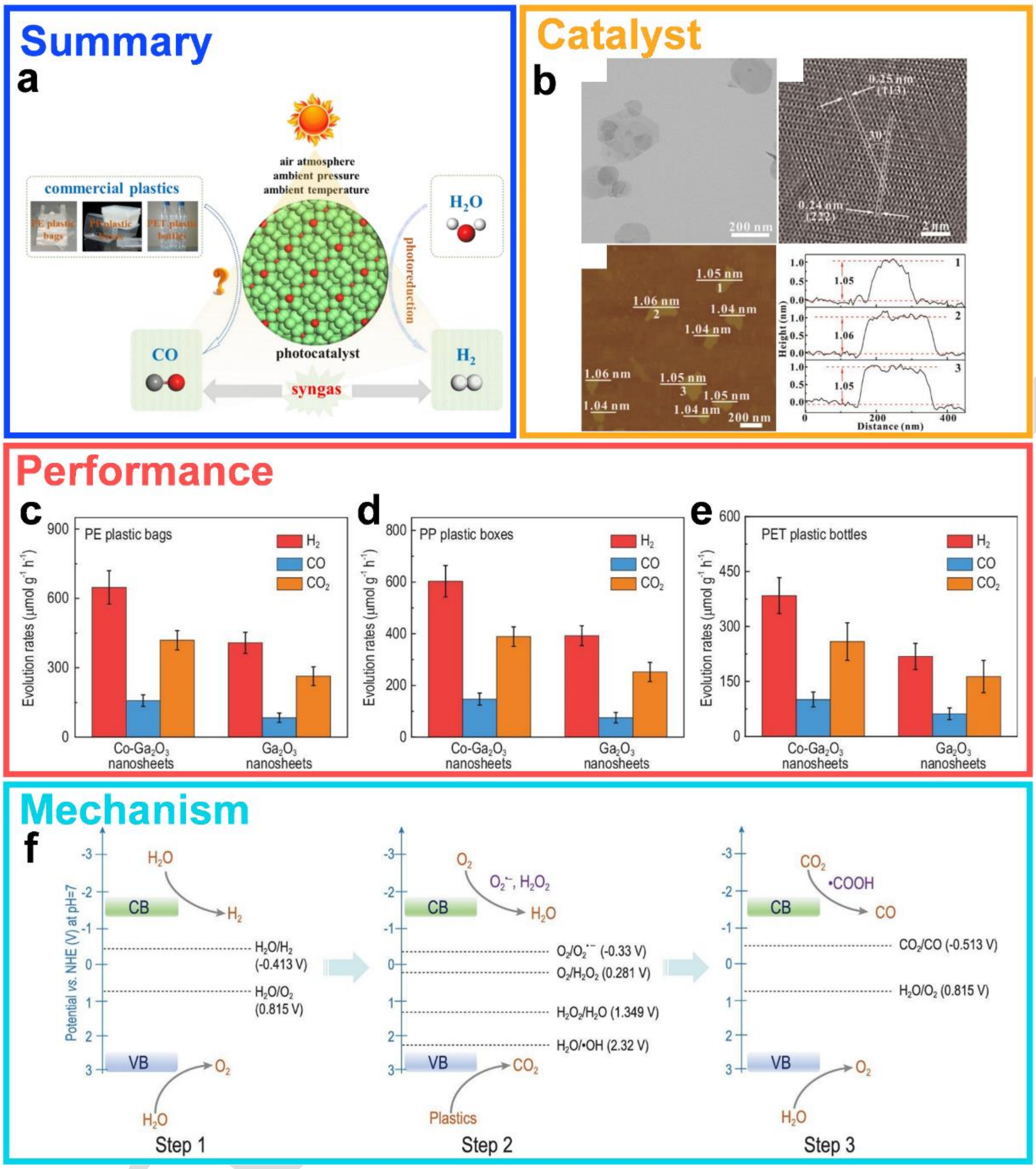

**Figure 20.** (a) Schematic representations for the photoconversion of non-recyclable plastics into syngas. (b) TEM and AFM images of Co-$Ga_2O_3$ nanosheets. Photoconversion of (c) commercial PE plastic bags, (d) commercial PP plastic boxes and (e) commercial PET plastic bottles under simulated AM 1.5G solar irradiation at ambient temperature and pressure. (f) Schematic representation of the proposed mechanism for the photoconversion of non-recyclable plastics into renewable syngas under mild conditions. Reproduced with permission.[108] Copyright 2022, Science Press.

*4.2.2. Photocatalytic conversion into materials*

In addition to the production of green fuels, researchers have explored the conversion of waste plastics into functional materials, which has garnered significant attention. Functionalizing the C–H bonds of plastics, rather than breaking these bonds, offers several advantages, including lower energy requirements.[109] A recent study by Leibfarth et al.[110] presented a method for functionalizing the C−H bond of PS to generate electrophilic fluoroalkyl radicals under mild conditions (**Figure 21a**). Ru(bpy)$_3$Cl$_2$ and organic catalysts such as phenoxazines, phenothiazines, and dihydrophenazines (**Figure 21b,c**) were utilized to generate electrophilic fluoroalkyl radicals (**Figure 21d,e**). The fluorination of PS increased the water contact angle significantly, indicating an enhanced hydrophobicity (**Figure 21f**). The optimal catalyst loading concentration for achieving the desired reactivity and selectivity was 1 mol% and 0.4 M 5,10-di(2-naphthyl)-5,10-dihydrophenazine (**Figure 21g,h**). Photocatalytic C-H alkylation is a promising route for converting plastic wastes into functional materials because of the relatively lower energy requirements for C-H bond activation. However, the limitations in the number of cycles and process control necessitate further exploration.

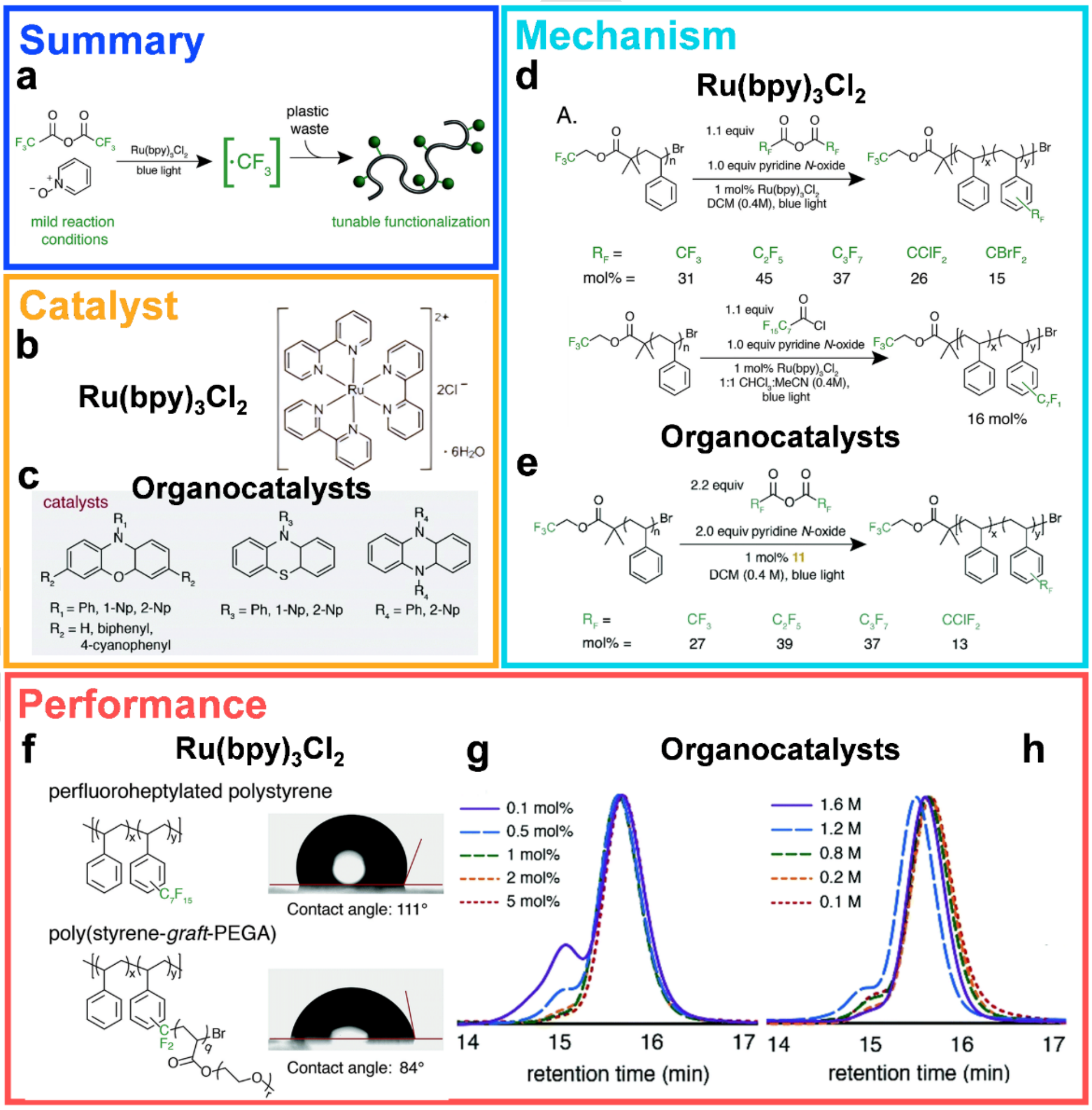

**Figure 21.** (a) Diagram of the functionalization of PS waste using electrophilic radicals under mild reaction conditions. (b) The structure of $Ru(bpy)_3Cl_2$. (c) The organocatalysts used in the perfluoroalkylation of PS. (d) Alternative perfluoroalkyl sources for polymer functionalization. Reproduced with permission.[110] Copyright 2019, The Royal Society of Chemistry. (e) Alternate acid anhydrides enable a greater diversity of functionality added to aromatic-containing polymers. Reproduced with permission.[111] Copyright 2020, The Royal Society of Chemistry. (f) Contact angle measurements highlighting the increased hydrophobicity of a series of perfluoroalkylated PS, suitable for coating applications. Reproduced with permission.[110] Copyright 2019, The Royal Society of Chemistry. GPC traces from the trifluoromethylation of PS (g) using different concentrations of the catalyst 5,10-di(2-naphthyl)-5,10-dihydrophenazine and (h) at different overall reaction concentrations Reproduced with permission.[111] Copyright 2020, The Royal Society of Chemistry.

*4.2.3 Photocatalytic conversion into chemicals*

As a carbon- and hydrogen-rich energy material, the conversion of plastics into high-value chemicals is important for reducing energy consumption. Plastic photoreforming, in contrast to thermal catalytic conversion, is advantageous because it does not require additional energy input and can be conducted under milder reaction conditions.[112] By harnessing renewable solar energy, plastic photoreforming enables precise cleavage of specific chemical bonds, resulting in high selectivity. In the plastic photoreforming reaction, solar radiation excites the photocatalyst, generating electron-hole pairs that participate in redox reactions at the material surface.[113] However, semiconductor photocatalysts often suffer from drawbacks such as the recombination of photogenerated carriers and slow reaction kinetics, which severely limit the conversion rate of waste plastics.[114]

To address these challenges, Qiao et al.[115] introduced a defect-rich chalcogenide-coupled photocatalyst for the solar-driven plastic photoreforming of hydrogen and high-value chemicals. They successfully constructed a defect-rich nickel-based chalcogenphosphate-coupled cadmium sulfide photocatalyst (d-$NiPS_3$/CdS) (**Figure 22a**) that could simultaneously convert PLA and PET plastic wastes under ambient conditions, resulting in hydrogen production rates as high as 40 $mmol \cdot g_{cat}^{-1} \cdot h^{-1}$ (**Figure 22b,c**). Importantly, the photoreforming reactions of waste PET plastic bottles and PLA plastic cups remained stable for over 100 h (**Figure 22d,e**). The oxidation products of PLA during a continuous 9-h reaction were acetates and pyruvate-based products, including pyruvate and its derivatives, under alkaline conditions. The oxidation products of PET included formate, acetate, and glycolate (**Figure 22f,g**). The efficiency of the plastic photoreforming process is attributed to the rapid separation of photogenerated electron-hole pairs facilitated by d-$NiPS_3$ nanosheets and efficient redox reactions at the defect-activated P and S sites on the photocatalyst surface (**Figure 22h**). This study provides new ideas for the design of efficient photocatalysts and sustainable conversion of plastics to high-value chemicals and green fuels. However, precise conversion of chemicals has not yet been achieved, posing a challenge for the effective utilization of complex products. Hence, this aspect requires further investigation.

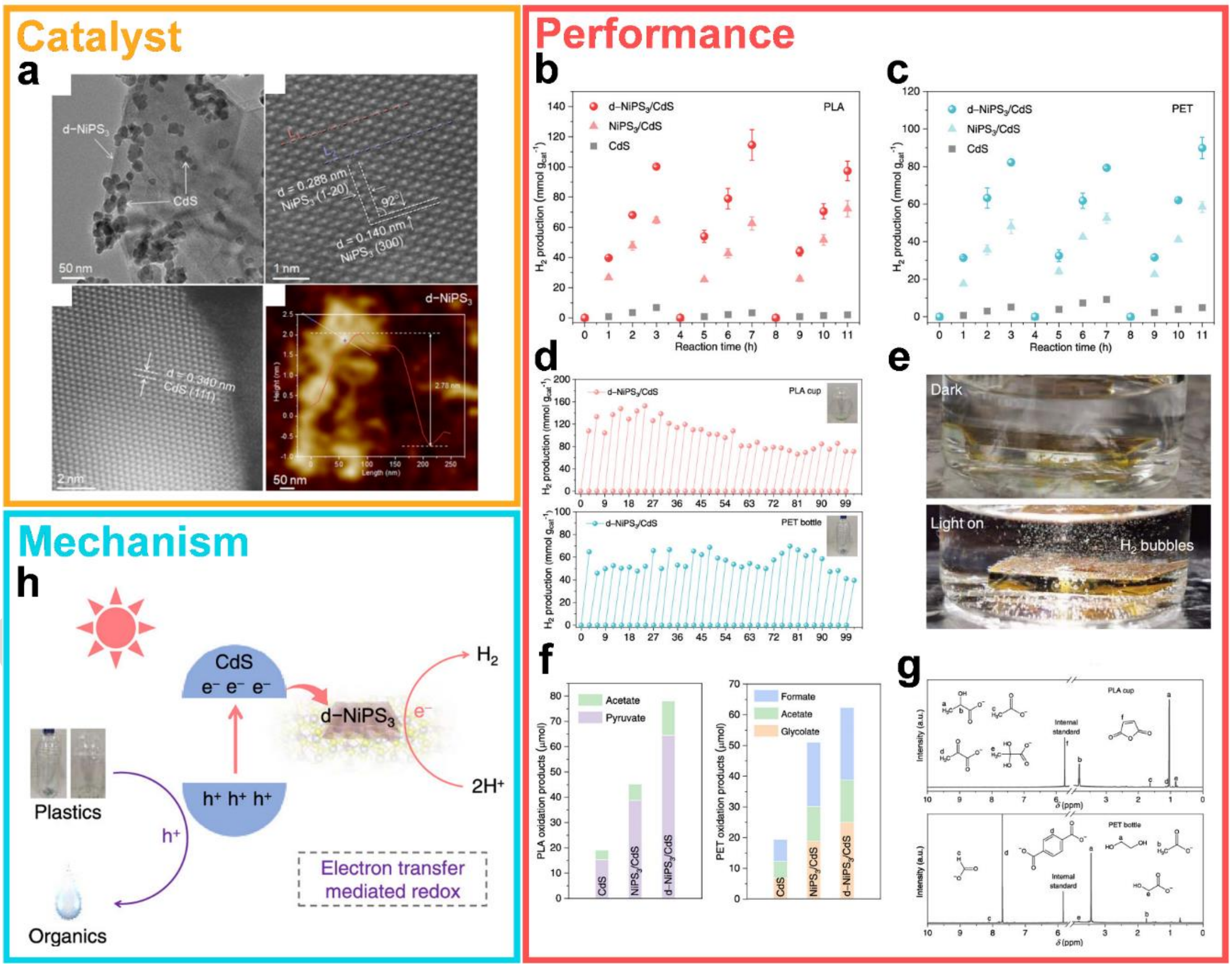

**Figure 22.** (a) Morphology and chemical composition for d-$NiPS_3$ and d-$NiPS_3$/CdS. Three runs of photoreforming for PLA (b) and PET (c) for $H_2$ production by CdS, $NiPS_3$/CdS, and d-$NiPS_3$/CdS under full-spectrum irradiation. (d) Consecutive $H_2$ production via long-term photoreforming of PLA cup (upper) and PET bottle (lower) with the d-$NiPS_3$/CdS photocatalyst. (e) Digital photograph of $H_2$ evolution using the d-$NiPS_3$/CdS film without and with illumination. (f) Organic acid products from PLA and PET for CdS, $NiPS_3$/CdS, and d-$NiPS_3$/CdS following 9 h of photoreforming. (g) $^1$H NMR spectra for substrates of PLA cups (upper) and PET bottles (lower) following photoreforming. Reproduced with permission.[115] Copyright 2023, American Chemical Society.

The photoreforming of plastics into high-value chemicals is a promising strategy. However, the complex nature of the products often poses challenges in their effective separation and utilization.[116] Consequently, there is a growing focus on achieving highly selective conversion of plastics into carbon-rich chemicals. In a recent study, Xie et al.[29] introduced a strategy for the selective conversion of waste plastics into $C_2$ chemicals via light-induced C–C bond cleavage and coupling (**Figure 23a**). The use of an ultrathin $Nb_2O_5$ photocatalyst with an appropriate redox potential led to a significant increase in $CH_3COOH$ production from PE, PP, and PVC. (**Figure 23b,c**). The reaction mechanism, as elucidated by *in situ* FTIR spectroscopy, revealed a two-step process for the photoreforming of PE into $CH_3COOH$ (**Figure 23d**). This process involved the oxidative cleavage of C−C bonds by $O_2$ and •OH to degrade long-chain molecules and generate $CO_2$ monomers, which subsequently underwent reduction to produce $CH_3COOH$. (**Figure 23e**). This study provides valuable insights into the two-step photoreforming mechanism for the conversion of plastics to chemicals. Optimization of the C−C bond cleavage and coupling processes to realize the selective photoreforming of plastic wastes into multi-carbon chemicals under natural environmental conditions will be a promising pathway for plastic upcycling. However, converting plastics to $CO_2$ and utilizing this $CO_2$ may lead to significant greenhouse gas emissions. Additionally, the two-step reaction process can limit the reaction rate.

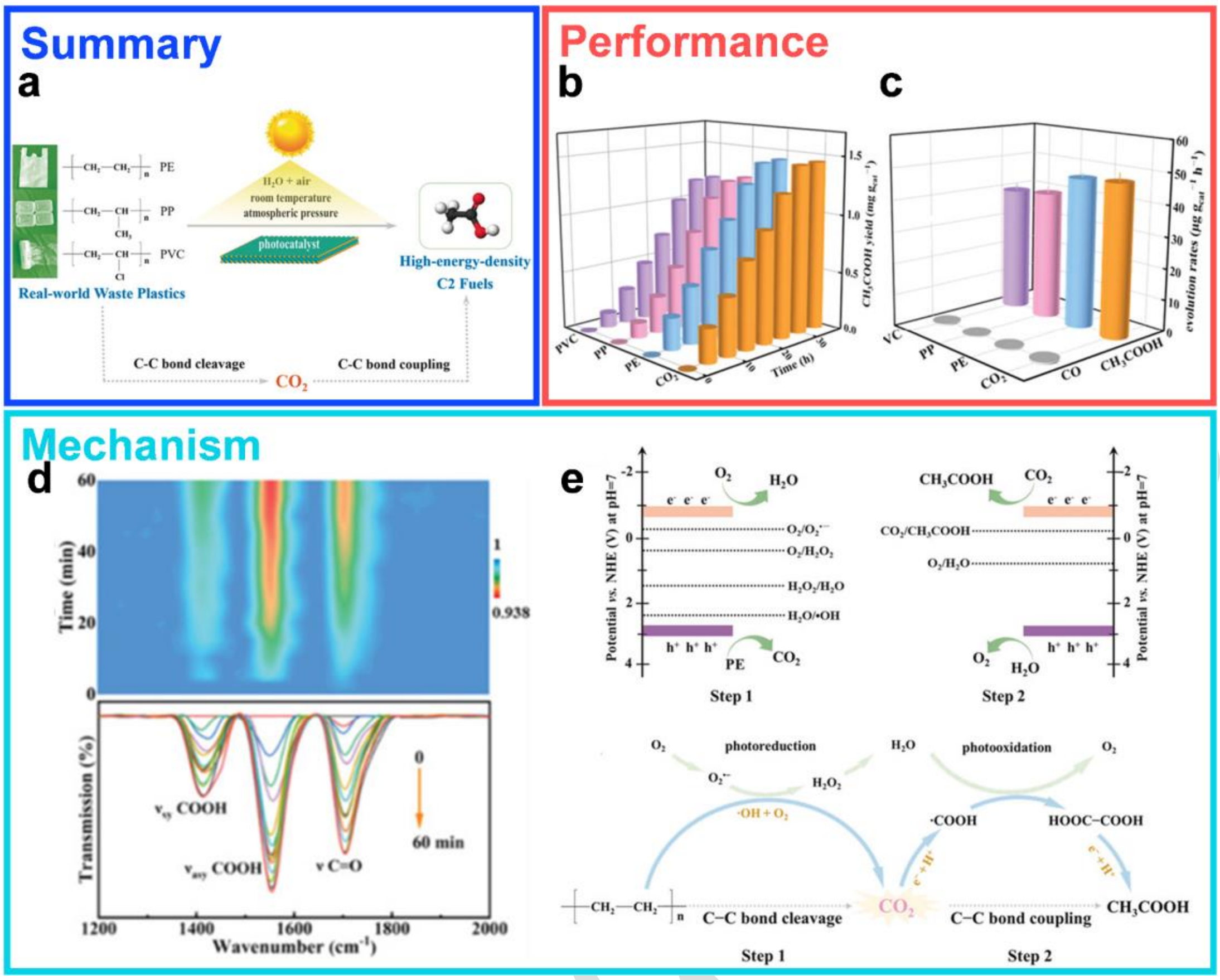

**Figure 23.** (a) Schematic illustration for converting various waste plastics into $C_2$ fuels by a designed two-step pathway under simulated natural environment conditions. (b) The yield of $CH_3COOH$ and (c) the evolution rates of $CH_3COOH$ and CO during the photoconversion of pure PE, PP, and PVC, as well as during the photoreduction of pure $CO_2$ in water. (d) *In-situ* FTIR spectra for detecting the reaction intermediates during PE photoconversion under simulated natural environment conditions. (e) The proposed two-step C−C bond cleavage and coupling mechanism from pure PE into $CH_3COOH$ under simulated natural environment conditions. Reproduced with permission.[29] Copyright 2020, John Wiley and Sons.

### *4.2.4. Other notable plastic photoreforming technologies*

In recent years, significant progress has been made in the field of plastic photoreforming. However, these developments are still in the early stages, leaving ample room for improvement in terms of practical prospects, product selectivity, reaction stability, and economic feasibility. Recently, Hyeon et al.[117] developed a floating photocatalytic platform comprising porous elastomer-hydrogel nanocomposites that demonstrated remarkable catalytic stability and potential for large-scale applications. By employing the Pt/$TiO_2$ low-temperature aerogel composite photocatalytic platform, plastic photoreforming experiments were conducted on PET, resulting in high catalytic stability even in strongly alkaline solutions. The hydrogen production rate reached 3.45 mmol·$h^{-1}$·$m^{-2}$, enabling the transformation of PET into valuable chemicals. The elastomer-hydrogel nanocomposite-based floating photocatalytic platform enabled efficient light transmission, facilitated convenient gas separation, inhibited reverse oxidation, and significantly enhanced the photocatalytic performance.

The implementation of multi-reaction coupling technology can further enhance the reaction rate and selectivity of plastic reforming. Reisner et al.[118] utilized the photoelectrochemical technology to achieve plastic reforming and $CO_2$ reduction. A perovskite-based photocathode capable of integrating various $CO_2$ reduction catalysts such as molecular cobalt porphyrin, $Cu_{91}In_9$ alloy, and formate dehydrogenase enabled the production of CO, syngas, and formate, respectively. Additionally, the $Cu_{27}Pd_{73}$ photoanode catalyst demonstrated high selectivity in alkaline solutions for the reforming of PET to glycolate. The performance of this system was comparable to that of an unbiased double photoabsorber, with a productivity approximately 10-100 times higher than that in the photocatalytic process. The solar-driven photoelectrocatalytic technology presented in this study offers a viable and efficient approach to plastic reforming. A comprehensive overview of the research advancements in plastic photoreforming is presented in **Table 6**. Enhancing the photocatalytic efficiency and achieving highly selective reforming are currently the primary areas of research. These aspects are crucial for the practical application of the photocatalytic technology.

**Table 6.** Photocatalytic conversion of plastics over different catalysts.

| Catalysts | Plastics | Light Source | Products | Selectivity | Refs. |
|---|---|---|---|---|---|
| $Nb_2O_5$ | PP、PE、PVC | 300 W Xe lamp with AM 1.5G filter | acetic acid | 100% | [116] |
| $CN_x$\|$Ni_2P$ | PET | Xe lamp with AM 1.5G filter (100 mW·cm$^{-2}$) | acetate and formate | 100% | [119] |
| $MoS_2$/CdS | PLA | 300 W Xe lamp with AM 1.5G filter | formate | 100% | [120] |
| C/g-$C_3N_4$ /Methanosarcina | PLA | ultraviolet LEDs | $CH_4$ | 100% | [121] |
| Fluorenone | PS | LED lamp | benzoic acid | 100% | [122] |

## 5. Conclusion and Outlook

The conversion of plastic wastes into high-value products such as chemicals and green fuels through plastic reforming is a transformative strategy. However, it is still in its nascent phase, primarily because of the challenges related to reaction efficiency, economic viability, and product selectivity. In light of the recent advancements in plastic reforming, we conducted a comprehensive review and have presented an outlook on plastic upcycling (**Figure 24**).

1) Achieving milder reaction conditions and higher efficiency: Thermocatalytic conversion has been extensively explored for the reforming of plastics at lower temperatures. The optimization of catalytic materials, such as composite catalysts and ionic liquid catalysts, should be focused for enhancing the catalytic performance while reducing the use of precious metals. Additionally, field-assisted approaches, such as microwave field technology and pulse heating technology, have been employed to increase thermal catalytic reaction rates. These strategies offer promising pathways for practical thermocatalytic plastic conversions. In electrocatalytic conversion, reducing the energy consumption remains challenging. Achieving efficient depolymerization of plastics at a lower applied bias and higher Faradaic efficiency is essential for minimizing reaction costs and enabling large-scale applications. Although photocatalytic conversion utilizes renewable solar energy, the need for strong alkali pretreatment and a relatively low reaction rate limits its broader application. Enhancing the catalytic ability of photocatalysts by loading co-catalysts or constructing heterojunctions can improve their overall performance.

2) Increasing selectivity in plastic reforming: High selectivity in plastic reforming is critical for realizing economical reactions, particularly for the production of high-value chemicals. Thermocatalytic conversion faces challenges because the precise breaking of specific chemical bonds under high-temperature conditions is difficult. The dissociation of C-C bonds at elevated temperatures leads to the formation of complex and low-value carbon monomers. Thus, catalysts should be selected such that they promote the interaction of plastic monomers on the catalyst surface. In the field of electrocatalysis, selective conversion is limited to plastic derivatives such as ethylene glycol. However, future trends aim to directly achieve the highly selective electrocatalytic reforming of plastics, without the need for derivative transformations. This requires the development of multi-reaction coupling processes in electrocatalysis to enhance the efficiency and achieve precise activation of the functional groups. For photocatalytic conversion, the development of low-cost and highly selective photocatalysts has been emphasized. Achieving high selectivity for the conversion of plastics into high-value chemicals is vital for practical applications. Photocatalysts with narrow bandgaps enhance the absorption of solar photons, whereas co-catalysts or heterojunctions promote charge separation and improve product selectivity.

3) Investigating the mechanism of plastic reforming: Understanding the conversion mechanism of plastics is essential for designing efficient reforming systems and achieving highly selective conversions. Advanced characterization techniques and theoretical calculations have played crucial roles in providing profound insights into the reaction mechanisms and interaction pathways between plastics and catalyst surfaces. *In situ* characterization methods provide valuable insights into the fundamental aspects of reaction processes. However, the applicability of *in situ* characterization techniques is often limited to milder reaction conditions. Theoretical calculations were employed to simulate the evolution of the reaction process and gain a deeper understanding of the reaction pathways. Calculations are frequently utilized to study the activation process of plastic monomers, analyze the changes in the Gibbs free energy for the conversion from monomers to specific selective products, and evaluate the changes in the reaction barrier. Theoretical calculations serve as a valuable guide for achieving highly selective reforming of plastic.

4) Novel coupled plastic reforming strategy: The utilization of multicomponent coupling reactions is crucial for achieving multifunctional and efficient plastic reforming. Such strategies are currently viable for electrocatalytic and photocatalytic applications. Electrocatalytic plastic reforming occurs at the anode through a cathodic reaction involving proton reduction for hydrogen generation. Thus, the efficiency or production of high-value products can be enhanced by employing multi-reaction-coupling with $CO_2$ reduction and biomass conversion. The coupling of plastic photoreforming with hydrogen generation offers thermodynamic advantages over direct water splitting. This can be expanded to include $CO_2$ reduction, $N_2$ fixation, metal recovery, and organic synthesis. External field coupling technologies (e.g., piezoelectric, magnetic, and ultrasonic fields) significantly enhance reaction efficiency and product selectivity. The synergistic effects of external field assistance also exert a notable influence on product selectivity.

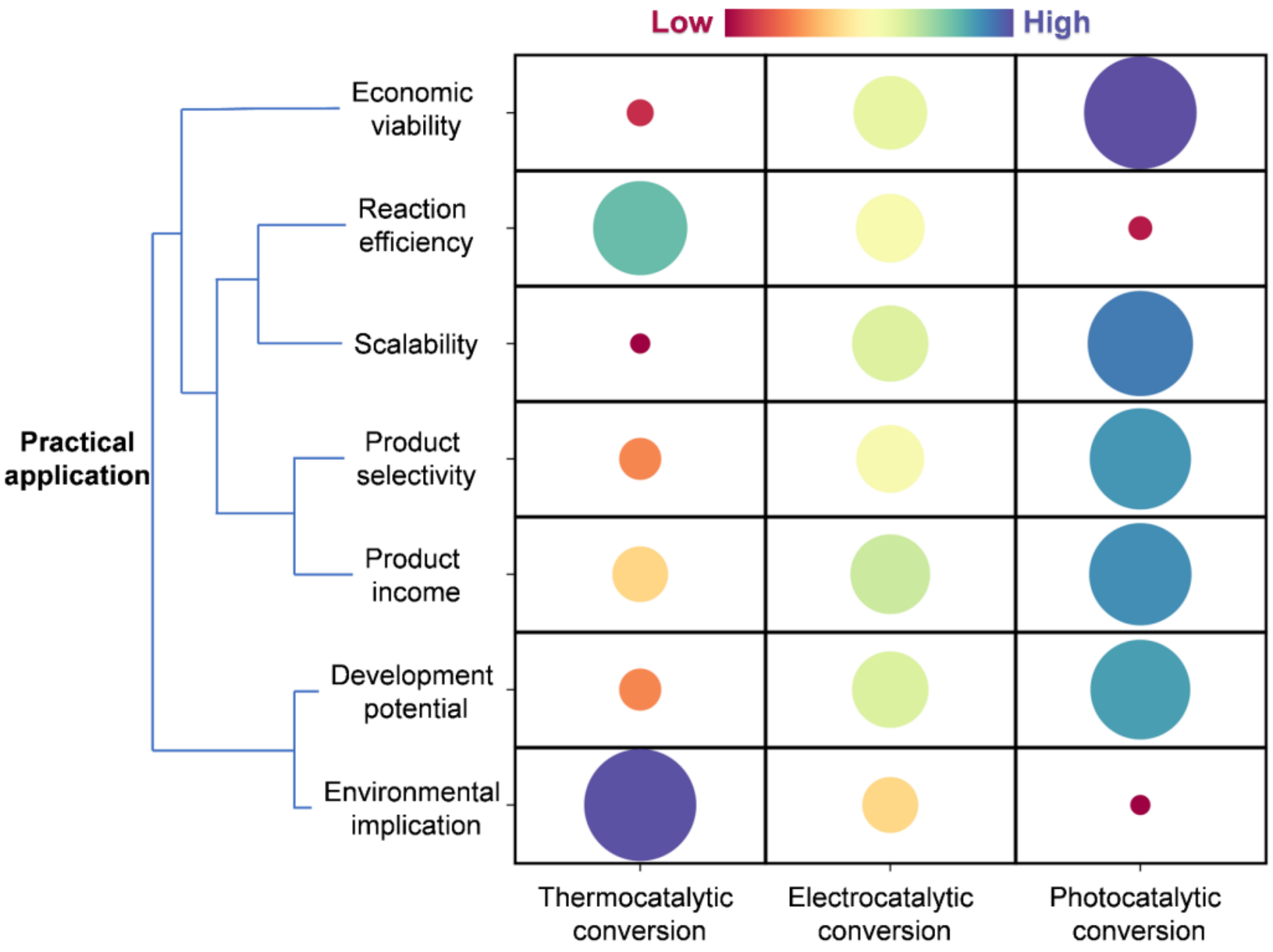

**Figure 24.** Correlation analysis between upcycling processes and main evaluation characteristics towards practical application. Economic viability: the reaction cost. Reaction efficiency: the average reaction speed. Scalability: the potential coupled with other reactions. Product selectivity: the proportion of a specific product. Product income: the added value of the products. Development potential: the capability for further advance. Environmental impact: the potential pollution and damage to the environment.

**Acknowledgements**

This work was supported by the support by the Natural Science Foundation of China as general projects (grant Nos. 22225604, 22076082, 21874099, 22176140, 22006029, and 42277059), the Tianjin Commission of Science and Technology as key critical technologies R&D projects (grant No. 21YFSNSN00250), the Frontiers Science Center for New Organic Matter (grant No. 63181206), and Haihe Laboratory of Sustainable Chemical Transformations.

**Conflict of Interest**

The authors declare no conflict of interest.

**Biographies**

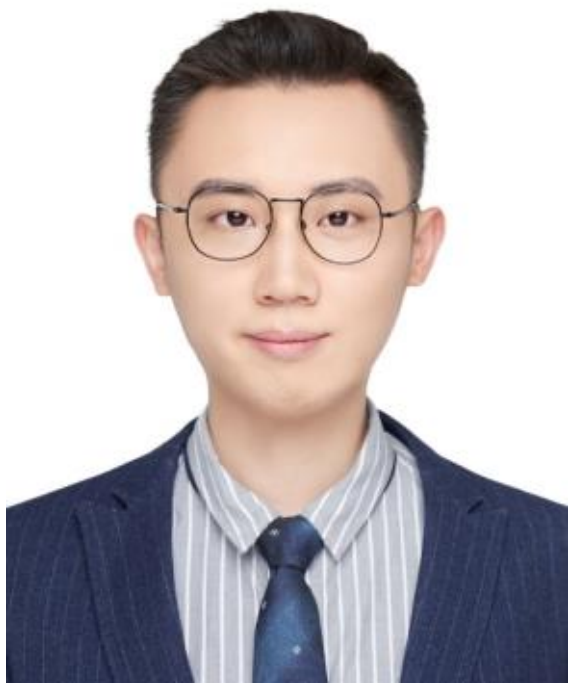

**Shuai Yue** is currently a Ph.D. candidate in Prof. Sihui Zhan's group, Nankai University. His research interest is focused on environmental catalysis, including waste plastics catalysis and single-atoms catalysis.

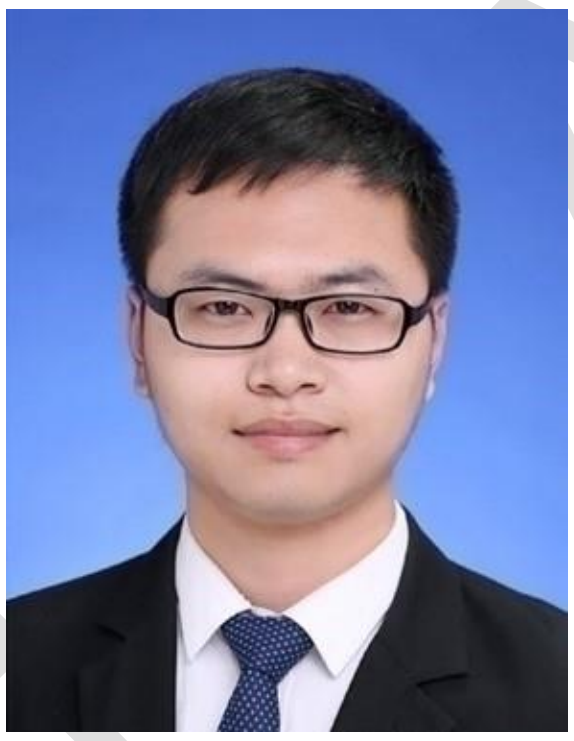

**Pengfei Wang** obtained his Ph.D. degree at Nankai University (2018). He joined the Faculty of Environmental Science and Engineering, Nankai University (2023). His research interests focus on environmental catalysis and waste plastic valorization.

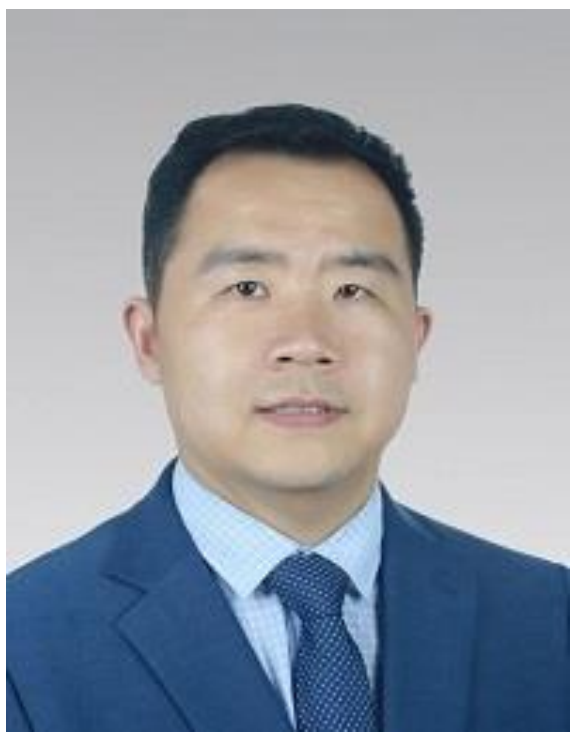

**Sihui Zhan** received his Ph.D. degree in inorganic chemistry from Shandong University and then joined Nankai University in 2007. He is now a full professor in the Faculty of Environmental Science and Engineering. He has research interests in fabrication of functional nanomaterials for water purification and environmental catalysis.

The table of contents:

Herein, a comprehensive overview of innovative catalytic technologies for the selective upcycling of plastic waste into high-value products is provided, including thermocatalysis, electrocatalysis and photocatalysis. Special attention is given to elucidating the mechanisms of plastic conversion and the design of catalysts.

Shuai Yue, Pengfei Wang*, Bingnan Yu, Tao Zhang, Zhiyong Zhao, Yi Li and Sihui Zhan*

**From Plastic Waste to Treasure: Selective Upcycling through Catalytic Technologies**

TOC figure

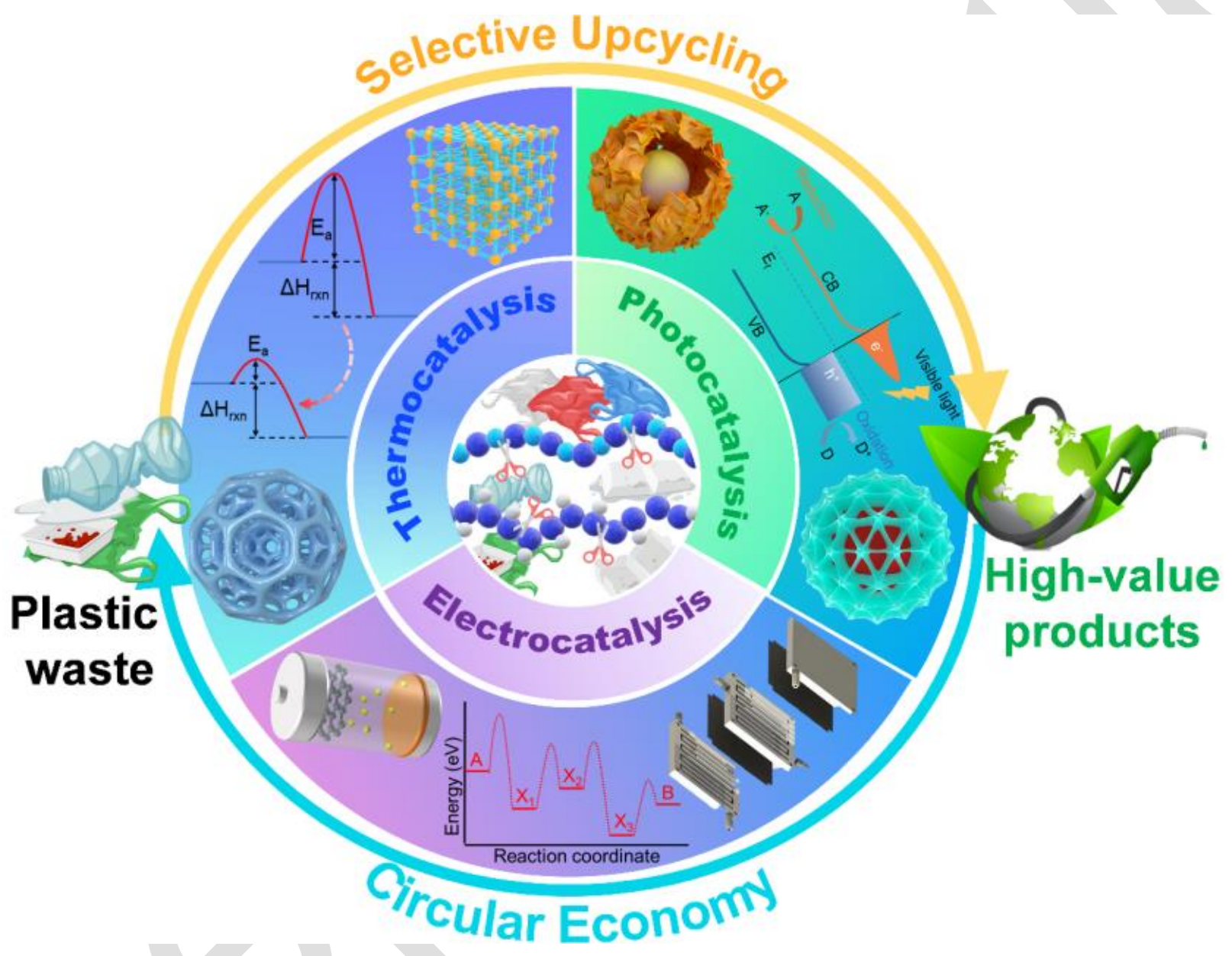